\documentclass[final,5p,twocolumn]{elsarticle}
\biboptions{sort&compress}
\usepackage[utf8]{inputenc}
\usepackage{amsmath}
\usepackage{amssymb}
\usepackage{graphicx}
\usepackage{acro}
\usepackage[hyphens]{url}
\usepackage{hyperref}
\usepackage[capitalize]{cleveref}
\usepackage{enumitem}
\usepackage{multirow}
\usepackage{siunitx}
\usepackage{float}
\usepackage[flushleft]{threeparttable}
\usepackage[table,xcdraw]{xcolor}
\usepackage[switch]{lineno}
\usepackage{tikz}
\DeclareAcronym{CSF}{
    short = CSF,
    long = cerebrospinal fluid
}

\DeclareAcronym{HCP}{
    short = HCP,
    long = Human Connectome Project
}

\DeclareAcronym{RS}{
    short = RS,
    long = Rhineland Study
}

\DeclareAcronym{MRI}{
    short = MRI,
    long = Magnetic Resonance Imaging 
}

\DeclareAcronym{MR}{
    short = MR,
    long = Magnetic Resonance
}

\DeclareAcronym{CNN}{
    short = CNN,
    long = Convolutional Neural Network
}

\DeclareAcronym{VINN}{
    short = VINN,
    long = Voxel-size Independent Neural Network
}

\DeclareAcronym{FastSurferVINN}{
    short = FastSurferVINN,
    long = FastSurferVINN 
}
\acuse{FastSurferVINN}

\DeclareAcronym{WM}{
    short = WM,
    long = white matter
}

\DeclareAcronym{GM}{
    short = GM,
    long = gray matter
}    

\DeclareAcronym{PVE}{
    short = PVE,
    long = partial volume effect}
    
\DeclareAcronym{abide-ii}{
    short = ABIDE-II,
    long = Autism Brain Imaging Data Exchange II}
    
\DeclareAcronym{NN}{
    short = NN,
    long = nearest-neighbour
}

\DeclareAcronym{ASD}{
    short = ASD,
    long = Average Surface Distance
}

\DeclareAcronym{DSC}{
    short = DSC,
    long = Dice Similarity Coefficient
}

\DeclareAcronym{TODO}{
    short = TODO,
    long = compile fix still todo
}

\DeclareAcronym{HiRes}{
    short = HiRes,
    long = high-resolution
}

\DeclareAcronym{LowRes}{
    short = LowRes,
    long = low-resolution
}

\DeclareAcronym{MSDA}{
    short = MSDA,
    long = multi-source domain adaptation
}

\setcitestyle{square}

\journal{NeuroImage}

\newcommand{\PyTorch}{\mbox{PyTorch}}
\newcommand{\FastSurferCNN}{\mbox{FastSurferCNN}}
\newcommand{\FastSurferVINN}{\mbox{FastSurferVINN}}
\newcommand{\new}[1]{#1}
\newcommand{\newfigure}[1]{#1}

\begin{document}
\begin{frontmatter}

\title{FastSurferVINN: Building Resolution-Independence into Deep Learning Segmentation Methods - A Solution for HighRes Brain MRI}

\author[label1]{Leonie Henschel\corref{cor1}}
\author[label1]{David Kügler\corref{cor1}}
\author[label1,label2,label3]{Martin Reuter\corref{cor2}}
\cortext[cor1]{Equal Contribution.}
\cortext[cor2]{Corresponding author.}
\address[label1]{German Center for Neurodegenerative Diseases (DZNE), Bonn, Germany}
\address[label2]{A.A.\ Martinos Center for Biomedical Imaging, Massachusetts General Hospital, Boston MA, USA }
\address[label3]{Department of Radiology, Harvard Medical School, Boston MA,USA}

\begin{abstract}

Leading neuroimaging studies have pushed 3T MRI acquisition resolutions below \SI{1.0}{\milli\meter} for improved structure definition and morphometry. Yet, only few, time-intensive automated image analysis pipelines have been validated for high-resolution (HiRes) settings. Efficient deep learning approaches, on the other hand, rarely support more than one fixed resolution (usually \SI{1.0}{\milli\meter}).
Furthermore, the lack of a standard submillimeter resolution as well as limited availability of diverse HiRes data with sufficient coverage of scanner, age, diseases, or genetic variance poses additional, unsolved challenges for training HiRes networks. 
Incorporating resolution-independence into deep learning-based segmentation, i.e., the ability to segment images at their native resolution across a range of different voxel sizes, promises to overcome these challenges, yet no such approach currently exists. We now fill this gap by introducing a Voxel-size Independent Neural Network (VINN) for resolution-independent segmentation tasks and present FastSurferVINN, which (i)~establishes and implements resolution-independence for deep learning as the first method simultaneously supporting 0.7-\SI{1.0}{\milli\meter} whole brain segmentation, (ii)~significantly outperforms state-of-the-art methods  across resolutions, and (iii)~mitigates the data imbalance problem present in HiRes datasets.
Overall, internal resolution-independence mutually benefits both HiRes and \SI{1.0}{\milli\meter} MRI segmentation. 
With our rigorously validated FastSurferVINN we distribute a rapid tool for morphometric neuroimage analysis. The VINN architecture, furthermore, represents an efficient resolution-independent segmentation method for wider application. 

\end{abstract}
\end{frontmatter}

\section{Introduction}

While neuroimaging pipelines have benefited substantially from the standardization of \ac{MRI} at \SI{1.0}{\milli\meter}, the resulting fixed-resolution paradigm now hinders transition to \ac{HiRes} \ac{MRI}.
With the hope of advancing quantification of structural detail, increasing explanatory power, and improving our understanding of the brain in health and disease \cite{Wattjes2006,Stankiewicz2011,Mellerio2014,Solano-Castiella2011,Glasser2013,Zaretskaya2018}, leading large-cohort neuroimaging studies have started to acquire structural MRI at 3T field strength and 0.7-\SI{0.9}{\milli\meter} resolutions (see \cref{ssec:high-res-MRI}).
However, the lack of reference segmentations and limited diversity of \ac{HiRes} \ac{MRI} datasets (e.g.\ regarding scanner, disease, genetic variation) lead to substantial limitations for bias-free method development and validation. 
Additionally, since no de-facto standard resolution exists for \ac{HiRes} imaging, neuroimaging tools introducing \ac{HiRes} processing \cite{Zaretskaya2018,cbstools,ASHS,nighres,Gaser2016} have to provide resolution-independence instead of following the fixed-resolution paradigm.
Although \acp{CNN} deliver convincing performance under the fixed-resolution paradigm \cite{wachinger2018deepnat,quicknat,slant2019,Mehta2017,Chen2018VoxResNet,Sun2019,Ito2019,MeshNet,AssemblyNet2020,Henschel_2020}, no methodological solution leverages explicit knowledge of the native image resolution, consequently limiting all output segmentations to one pre-defined voxel size and potentially ignoring important structural detail specifically for submillimeter scans. 
By \textbf{introducing \aclp{VINN}} (\acp{VINN}), we now leverage the diversity of widely available \SI{1.0}{\milli\meter} \acp{MRI} and enrich the model with details derived from \ac{HiRes} \ac{MRI}, achieving not only resolution-independence but improving segmentation performance across resolutions.

\begin{figure*}[!hbt]
    \centering
    \includegraphics[page=8,width=\textwidth,keepaspectratio]{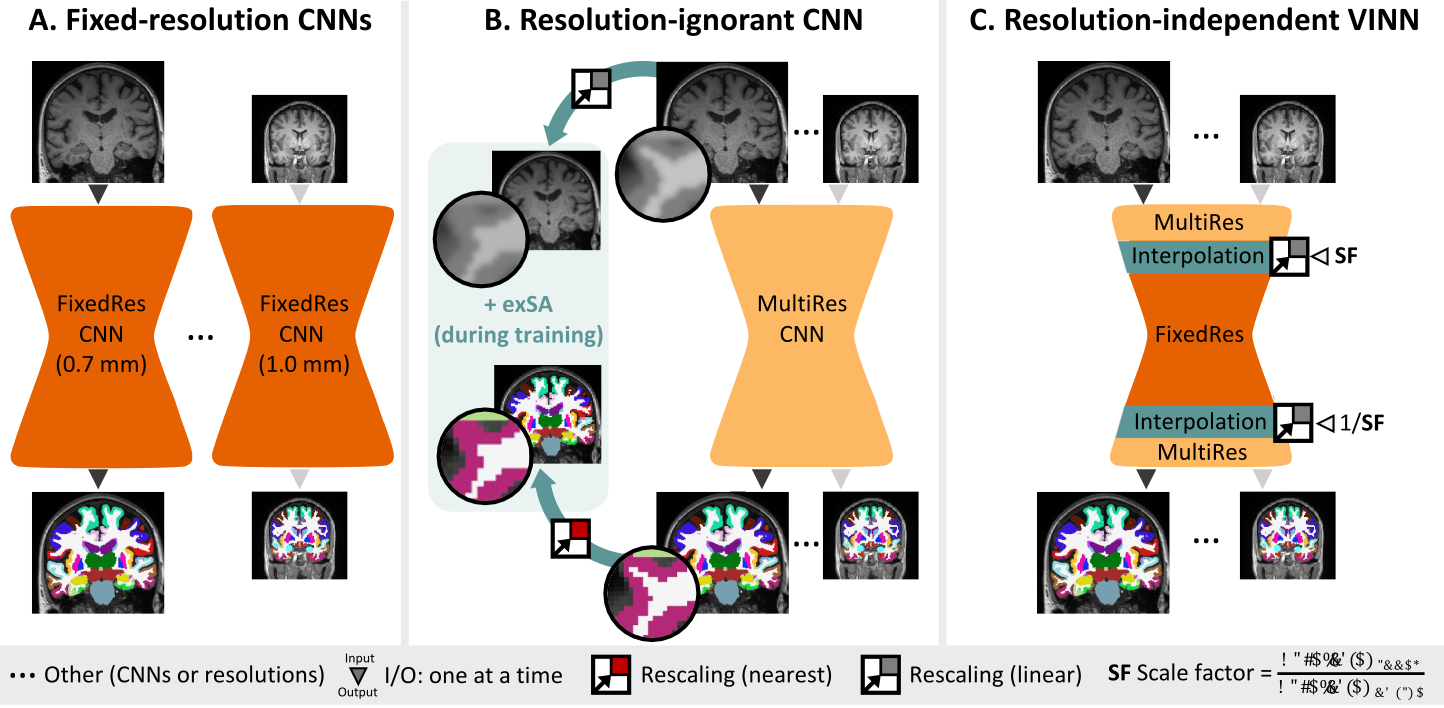}
    \caption{Resolution-independence in deep learning networks: \textbf{A.}~Dedicated fixed-resolution convolutional neural networks (CNNs) only work on the resolution they are trained on and are limited by the availability and quality of corresponding datasets. \textbf{B.}~One single resolution-ignorant CNN can learn to segment multiple resolutions by training on a diverse dataset. External scale augmentation (+exSA, B.\ left) simulates resolutions with few or no training cases by resampling the image and the reference segmentation map. Here, however, lossy interpolation and resulting artefacts, especially from nearest-neighbour interpolation of discrete label maps, may result in a loss of structural details and sub-optimal performance. \textbf{C.}~Our voxel size independent neural network (VINN) avoids interpolation of the images and discrete labels by integrating the interpolation step into the network architecture. Further, the explicit transition from the native resolution to a normalized internal resolution facilitates an understanding of the difference between image features (MultiRes blocks with distances measured in voxels) and anatomical features (FixedRes inner blocks with normalized distances). }
    \label{fig:networks}
\end{figure*}

For resolution-independent deep learning, we establish two core requirements:
1.~Native-resolution segmentation: the network's input and output for training \emph{and crucially inference} should be at the native resolution, to avoid any external resampling.
2.~Resolution-independence: the network should be able to learn and predict from images at a range of different resolutions.
Segmentation for \ac{HiRes} images additionally aims to improve the quality of fine structures (e.g.\ narrow sulci, gyri and \ac{WM} details).
To achieve high-quality whole brain segmentation and avoid training biases, a neural network should generalize to various resolutions (seen and unseen during training) and datasets with different characteristics (e.g.\ with respect to scanners, demographics, diseases, genetic variation) ideally sharing and transferring knowledge between resolutions.

While some traditional neuroimaging pipelines fulfill these requirements \cite{Zaretskaya2018,cbstools,ASHS,nighres,Gaser2016}, no related work directly addresses these challenges with deep learning. \Cref{fig:networks} illustrates two adaptations to form baseline solutions: A.~dedicated fixed-resolution networks and B.~a single resolution-ignorant \ac{CNN} that accepts multiple resolutions through training. Training one dedicated network per resolution (A.) trades the potential of a larger, more diverse training corpus for compatibility with the native resolution raising bias and generalization limitations (there are, for example, currently no \ac{HiRes} neurodegeneration datasets). However, a fixed-resolution network trained and evaluated on its native resolution represents an upper bound for achievable performance under same-size training datasets. On the other hand, if one network is na\"ively trained on multiple resolutions (B.), each convolutional layer has to learn to generalize across scales as the network cannot easily differentiate between different voxel sizes (i.e.\ resolution-ignorant), which allocates network capacity to this task. To support this process and reduce potential bias from missing or unbalanced resolution data, one can add external scale augmentation (+exSA) by resampling individual images and reference labels during training. This, however, induces information loss and interpolation artefacts, e.g.\ from lossy \ac{NN} interpolation of discrete label maps. Since neither approach (A.\ or B.) has been implemented or compared for submillimeter whole brain segmentation, we introduce respective baseline models utilizing our proposed, optimized micro-architecture in all models for a fair comparison.

To overcome limitations of both approaches (A.\ and B.), we propose a \ac{VINN} with innovations including its micro-architecture and the addition of a \ac{HiRes} loss. The core contribution, however, is the network-integrated resolution-normalization to support native segmentation at various voxel sizes. In fact, any UNet-based architecture \cite{Ronneberger2015} is, by design, a multi-scale approach where pooling operations represent fixed-factor integer down- and up-scale transitions (usually by the scale factor 2). Our network-integrated resolution-normalization in \ac{VINN} now replaces this fixed scale transition with a flexible re-scaling for the first and last scale transitions. \new{This has the advantage of placing our interpolation operation at a position where information loss naturally occurs (down- or up-scaling via pooling)}. As illustrated in \Cref{fig:networks}C.,\ we retain compatibility with a range of resolutions for input and output (MultiRes) by shifting the interpolation into the architecture itself. At the same time, we leverage the lower variance of perceived size differences in the inner normalized resolution blocks (FixedRes). This has the advantages of (i)~retaining important image information at the native resolution in the MultiRes blocks, (ii)~interpolating multi-dimensional continuous feature maps rather than discrete labels or single slice images hence avoiding lossy \ac{NN} interpolation and extending contextual neighbourhood information during the resolution-normalization, and (iii)~disentangling perceived voxel versus actual structure size differences inside the \ac{VINN}. 

Especially the last point may have been underappreciated so far: Due to the nature of convolutional layers, \acp{CNN} `perceive' distances and thus structure sizes by number of voxels rather than millimeters. Therefore, the original voxel size impacts perceived distances for baseline architectures, requiring them to cover larger size variety during training and inference in the multi-resolution context. The \ac{VINN}, on the other hand, transitions from a voxel-based distance context in the MultiRes blocks to a standardized distance context in FixedRes via the network-integrated resolution-normalization and, therefore, releases network capacity in the inner layers for other tasks. 

\subsection{Contributions}

All in all, our \ac{VINN}, for the first time, effectively addresses the challenges associated with \ac{HiRes} \ac{MRI} (i.e.\ multiple resolutions and reduced variety in datasets) in a single framework. We extensively test variations of the network architectures and demonstrate that inherent biases introduced by the unavailability of certain \ac{HiRes} scans are reduced by transferring information across resolutions via our network-integrated resolution-normalization. \\
Specifically, we show that our \ac{VINN}

\begin{enumerate}[label=\roman*.]

\item segments 3T brain MRI at their native resolutions (\SI{0.7}-\SI{1.0}{\milli\meter}) into 95 regions in less than 1 minute on the GPU and generalizes robustly to unseen resolutions within and beyond the training corpus, 

\item significantly outperforms state-of-the-art scale augmentations as well as fixed-resolution models with respect to segmentation accuracy and generalizability with an optimized architecture, and 

\item improves accuracy by combining and leveraging both the increased structural information from submillimeter 3T brain scans and the generalizability to intensity, scanner, disease, and other variations from standard \SI{1.0}{\milli\meter} \acp{MRI}.

\end{enumerate}

\section{Related Work}

\subsection{High-resolution MRI}\label{ssec:high-res-MRI}

In neuroimaging, spatial resolution is of great importance as the available voxel size directly dictates the degree to which fine-scale subcortical and cortical structures, specifically narrow gyri and sulci, can be resolved in an \ac{MRI}. In particular, the diversity of tissue types in a single voxel, which heavily depends on its size, influences the signal intensity known as the \acf{PVE}. A strong \ac{PVE} specifically complicates the delineation of tissue borders on the voxel grid \cite{Zaretskaya2018, Glasser2013}. \ac{HiRes} images offer finer sampling of the underlying information, thereby directly reduce \ac{PVE} and enable more accurate segmentations, improved volume-based measurements, morphometry, surface placement, and derived thickness measures (see \Cref{fig:PVE}) \cite{Luesebrink2013, Zaretskaya2018, Glasser2013}. 

\begin{figure}[!tb]
    \centering
    \includegraphics[width=0.9\columnwidth,keepaspectratio]{motivation_highRes2.png}
    \caption{Image resolution affects the detail of discrete segmentation label maps and derived measures such as surface models and thickness. \textbf{A.} The low-resolution image is less detailed and causes \acfp{PVE} by accumulating signals across tissue boundaries into larger voxels, whereas \textbf{B.} \ac{HiRes} images and derived segmentations allow more precise region delineation and capture details, e.g.\ for improved shape or thickness analysis (white arrows).}
    \label{fig:PVE}
\end{figure}

It is, therefore, not surprising that \ac{HiRes} \acp{MRI} are becoming increasingly popular within the neuroimaging community. While most established large-scale neuroimaging studies acquire data at \SI{1.0}{\milli\meter}, the current and next generation studies are shifting to submillimeter resolutions (e.g.\ \ac{HCP} \cite{Glasser2013}, \ac{RS} \cite{rhineland}, \ac{abide-ii} \cite{abide_dataset}, TRACK-PD \cite{track_pd}). 
However, to date the consolidated superset of publicly available 3T \ac{HiRes} neuroimages is sparse, unbalanced, and heterogeneous with respect to available submillimeter resolutions (e.g.\ \ac{HCP}: 0.7 and 0.8, \ac{abide-ii}: 0.7-0.9, \ac{RS}: 0.8). 
Especially the limited data variety at each specific resolution poses a real challenge for data-driven computational methods, which translates into limited compatibility with scanners, age spans, and especially disease groups. Moreover, unbalanced training data can easily lead to the introduction of biases into the model. 

While suffering from stronger \acp{PVE} and ensuing detrimental effects on segmentation accuracy, the available collection of standardized \SI{1.0}{\milli\meter} images, on the other hand, is large and diverse. Consequently, a wide coverage of age-groups, diseases, genetic variants, and scanners can be retrieved from the rich reservoir of openly available \ac{MRI} data sources (e.g.\ OpenfMRI database \cite{Poldrack2013,Poldrack2014}, OpenNeuro \cite{openNeuro}, NITRC-IR\footnote{\url{www.nitrc.org}}). Furthermore, manual reference labels for validation are exclusively published openly for \SI{1.0}{\milli\meter} \cite{Klein2012}. 
To address dataset sparsity and bias at submillimeter resolutions, we believe data-driven computational methods require built-in resolution-independence. Only a single model that spans across the available resolutions can simultaneously provide benefits for both \ac{HiRes} and standard \SI{1.0}{\milli\meter} image analysis.

\subsection{Automated analysis of high-resolution images}
Traditionally, common neuroimaging pipelines have been developed and optimized for \SI{1.0}{\milli\meter} voxels \cite{FSL, FSLFast, spm_book2007,Gaser2016,fischl2002whole} representing the de-facto standard for years. 
FreeSurfer offers a validated \ac{HiRes} stream \cite{Zaretskaya2018}, which provides sub-segmentation of the cortex into 31 structures per hemisphere (DKTatlas) \cite{Klein2012}. However, one problem here is the extended processing time caused by the cubic voxel increase -- a common issue limiting applicability of traditional tools to large cohort studies.  

The introduction of \acp{CNN} for whole brain segmentation has substantially reduced processing times to seconds on the GPU. Recent works employ both 2.5D and 3D UNet architectures \cite{wachinger2018deepnat,quicknat,slant2019,Mehta2017,Chen2018VoxResNet,Sun2019,Ito2019,MeshNet,AssemblyNet2020,Henschel_2020,Billot_2020,Iglesias_2021}. Since GPU memory limitations render full volume 3D models impractical specifically for higher number of feature channels and output classes, top performing methods process the volume in slices (QuickNat, FastSurfer \cite{quicknat,Henschel_2020}) or in large patches (DeepNat, SLANT, AssemblyNet \cite{wachinger2018deepnat,slant2019,AssemblyNet2020}) and then leverage aggregation schemes to recombine predictions into the full volume. 

Despite their success for brain segmentation and other applications no work has introduced deep learning for submillimeter whole brain segmentation. Since \acp{CNN} require a large and diverse number of volume and segmentation pairs for effective training, the missing availability of diverse training data hinders this resolution transition. Additionally, the cubic relationship between resolution increase and GPU memory requirements puts memory hungry 3D architectures at a disadvantage, e.g.\ factor of $2.92=(1/0.7)^3$ memory demand for a 1 to \SI{0.7}{\milli\meter} reduction of voxel sizes.

In the past, multi-branch segmentation frameworks \cite{Kamnitsas2017, Wang_2017,HookNet,Gu2018,Gerard2020,Fu2018,Liu2020,Yang2018,HRCNet,Zheng2021} have been used to avoid memory issues while simultaneously leveraging \ac{HiRes} information. Here, multiple potentially cross-linked pathways are dedicated to specific down-scaled or cropped versions of the original image. \new{The same principle is employed in other scale-aware networks \cite{Chen2016,Li2019,Huang2021} by implementing more trans-scale connections or combinations of different dilation and kernel sizes}. To further avoid sub-optimal equal weighting of different scale information, the networks often include attention mechanisms \cite{Chen2016,HRCNet,Yang2018,Qin2018,Zheng2021}. In practice, these methods are only compatible with the discrete input and sub-level resolutions they were trained on, \new{leaving subvoxel scaling more common in neuroimaging data (0.7/0.8/1) unexploited}. In fact, their intend is to explicitly integrate information from multiple image scales rather than achieve multi-resolution compatibility. 
\new{Critically, in neuroimaging settings the true scale is known (image resolution) while scale-aware architectures assume this knowledge is not available. In VINN we inject this explicit knowledge into the network at the spatial resolution normalization step.}

\subsection{Resolution-independence in deep learning}
Built-in resolution-independence in \acp{CNN} has not been described for brain segmentation nor -- to our knowledge -- for any other segmentation tasks. 
Approaches such as \cite{Billot_2020,Iglesias_2021} pre-sample input images (with associated reliability maps) to a common resolution (here \SI{1.0}{\milli\meter}) and provide outputs there, which makes them inherently fixed-resolution techniques. While they can provide \SI{1.0}{\milli\meter} segmentations (and even images) for lower resolutional clinical scans via heavy augmentation, they can neither profit from submillimeter details, nor provide native \ac{HiRes} segmentations.
A transfer to higher resolutions would require retraining with a fixed submillimeter training set. This is, however, problematic as (i)~no standard submillimeter resolution exists, meaning one would either need to train multiple versions or focus on the highest available resolution to retain input resolution-independence, (ii)~\ac{HiRes} datasets, as mentioned before, demonstrate low subject variance (disease, age, genetics) making this approach susceptible to training biases, and (iii)~upscaling the 3D-UNet architecture to \SI{0.7}{\milli\meter} would require 2.9-times as much GPU memory surpassing memory limits. 

Different from segmentation networks, state-of-the-art super-resolution networks aim specifically at reconstructing a (fixed resolution) \ac{HiRes} image based on a \ac{LowRes} input, also requiring ground truth at the higher resolution. They often rely on pre-sampling, i.e.\ using interpolation-based up-sampling methods to initialize the desired output grid and fine-tune the features in the following network \cite{SRCNN, VDSR, DRCN}. Using augmentation, these networks are trained to restore images coming from various lower resolutions via a single model, similar to the discussed segmentation strategy (\Cref{fig:multiresnet}B.). While showing better performance than dedicated single-scale models (fixed input and fixed output resolution), pre-sampling significantly increases the computational complexity and cost of the architectures due to the increase in image size. To overcome this limitation, interpolation-based post-sampling has been introduced recently \cite{EMSR,ASDN}. Here, the up-sampling step is shifted towards the end of the network architecture and performs the interpolation in the latent space. Interestingly, this post-sampling approach is as effective or even superior to the pre-sampling methods, while maintaining versatility with respect to the chosen output scale \cite{EMSR,ASDN}. 
All these super-resolution architectures, in addition to not being aimed at segmentation, differ from our approach by transitioning once, from the input resolution to an output resolution. In contrast, we insert two latent-space interpolation blocks transitioning both ways between the native and the inner resolution. 

\new{More generally, \acp{MSDA} leverage available data with different underlying distributions (i.e.\ resolutions in our case). This field of research focuses on enhancing the generalization ability of a model by transferring knowledge between resource-full source domains (i.e.\ \ac{LowRes} data) to a sparsely represented target domain (i.e.\ \ac{HiRes} data) by transforming either the features in the latent space or images on a pixel-level. In contrast to our resolution-normalization, the latent space transformations in \acp{MSDA} are based on optimizing a discrepancy \cite{Guo2018, Zhu2019, Peng2019} or adversarial loss \cite{xu2018, Li2019, Wang2019}. These latent space alignments are, however, often insufficient for segmentation tasks due to their focus on high-level information only \cite{Zhao2019a}. To circumvent this problem, intermediate domain generators have been proposed and successfully applied for semantic segmentation \cite{Hoffman2017,Russo2019,Zhao2019a}. Here, a pixel-level alignment between source and target domain is learned through Generative adversarial networks. However, these \ac{MSDA} methods are limited to a single target distribution. Extension to multi-targets is a relatively unexplored area with only a few published methods for classification tasks so far \cite{Chen19,Peng2019,Jin2020,Liu2020,Gholami2020,Yang2020,Roy2021}.}

Spatial transformers \cite{spatialT} represent another technology relying on internal interpolations as an important building block. Here, spatial invariance (registration) is targeted via a learnable affine transformation inside the network (localisation network). After the source feature map coordinates are computed (grid calculator), intensity values for each target pixel are determined via bi-linear interpolation (sampler). Spatial transformers hence attempt to implicitly learn data representations in a resolution-ignorant way.
While our approach shares grid calculation and interpolation within the network with spatial transformers, our training approach is different: Instead of a localisation network, we directly determine the sampling-grid based on the input scale factors, i.e.\ the ratio between the native input resolution and the desired normalized inner resolution. Hence, we explicitly integrate knowledge about the image resolutions into the architecture. As a result, computational complexity is reduced while still achieving desired resolution-independence.

\section{Material and methods}
\subsection{Datasets}\label{sec:datasets}
The following three submillimeter \ac{MRI} datasets were selected for training, testing, and validation of FastSurferVINN. An extended list of the used \SI{1.0}{\milli\meter} datasets can be found in the \Cref{sec:appendix_datasets}. When not specifically mentioned otherwise, all sets are balanced for gender, age, and study. Participants of the individual studies gave informed consent in accordance with the Institutional Review Board at each of the participating sites. Complete ethic statements are available at the respective study webpages.

\vspace{1ex}
    \textbf{ABIDE-II}: The Autism Brain Imaging Data Exchange II ~\cite{abide_dataset} contains cross-sectional data and focuses on autism spectrum disorders. The dataset contains 1114 subjects from 19 different institutions in total and is accessible online\footnote{\url{http://fcon_1000.projects.nitrc.org/indi/abide/abide_II.html}}. The 3D magnetization prepared rapid gradient echo (MPRAGE) sequence, or a vendor specific variant, was used to acquire all data using 3T GE, Philips, and Siemens scanners. Voxel resolutions are not standardized across sites although most scans are acquired at \SI{1.0}{\milli\meter}. In this work we use only the ETH\_1 sub-cohort, which provides \SI{0.9}{\milli\meter} \ac{HiRes} scans acquired on a 3T Philips Achieva with a repetition time (TR) of \SI{3}{\milli\second}, an echo time (TE) of \SI{3.90}{\milli\second}, an inversion time (TI) of \SI{1.15}{\milli\second}, and flip angle of \ang{8}. Since \ac{HiRes} datasets often exclusively feature Siemens scanners, a subset of 25 \ac{HiRes} scans from \ac{abide-ii} ETH\_1 covering an age range of 20-31 years serves as an independent test set to evaluate generalizability to this unseen Philips scanner and \SI{0.9}{\milli\meter} resolution.
    
\begin{figure*}[!hbt]
    \centering
    \newfigure{\includegraphics[width=\textwidth,keepaspectratio]{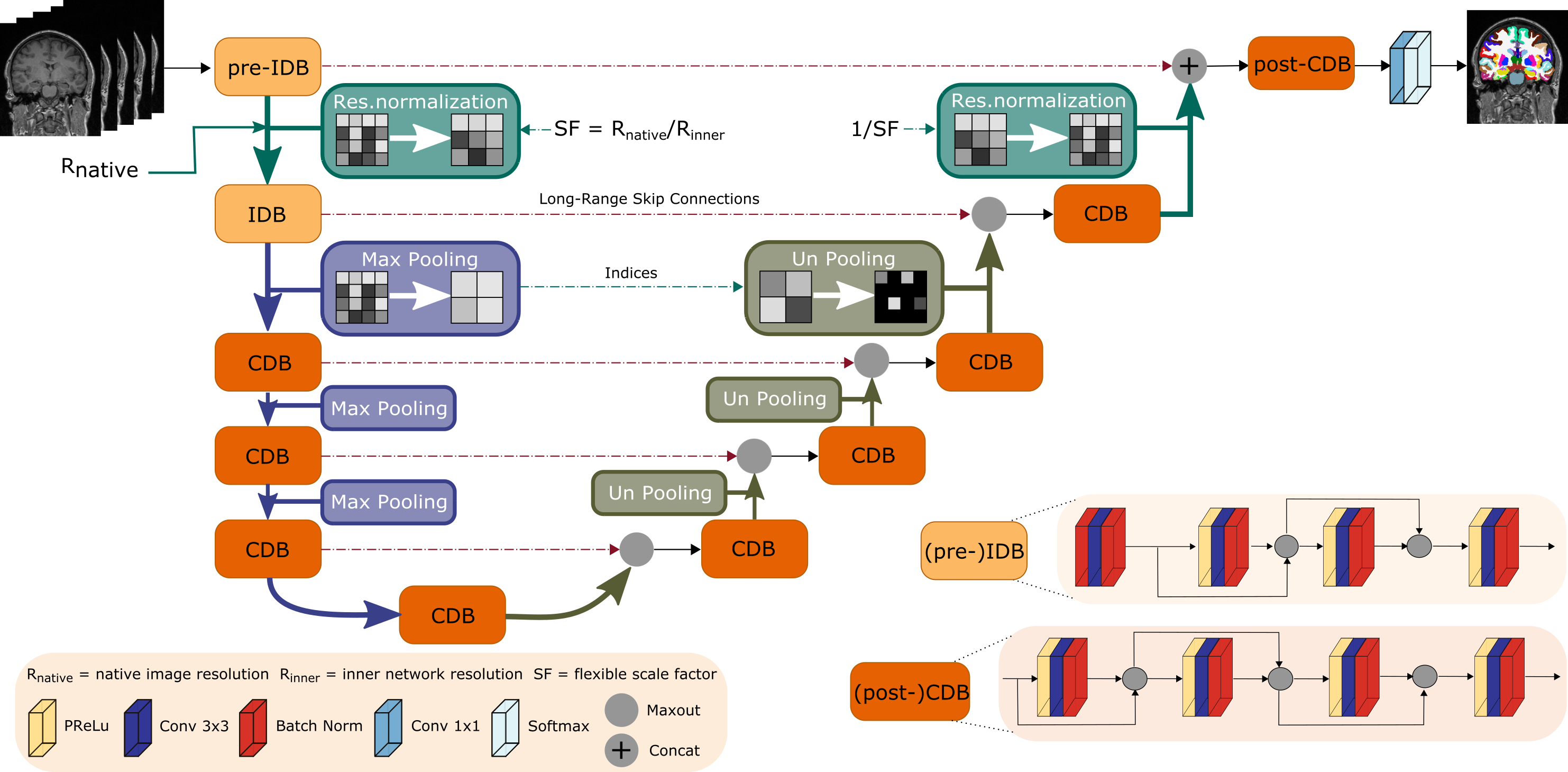}}
    \caption{Voxel size independence in FastSurferVINN. \new{Flexible transitions between resolutions become possible by replacement of (un)pooling with our network-integrated resolution-normalization (green) after the first encoder (pre-IDB) and before the last decoder block (post-CDB). Scale transitions between the other competitive dense blocks (CDB) remain as standard MaxPool and UnPool operations}. Each CDB is composed of four sequences of parametric rectified linear unit (PReLU), convolution (Conv) and batch normalization (BN). In the first two encoder blocks ((pre)-IDB), the PReLU is replaced with a BN to normalize the inputs.}
    \label{fig:multiresnet}
\end{figure*}
    
\vspace{1ex}
    \textbf{HCP}: The Human Connectome Project Young Adult~\cite{hcp_dataset} is a cross-sectional study including 3T \acp{MRI} from 1200 healthy participants acquired on a customized Siemens scanner (Connectome Skyra). It provides \SI{0.7}{\milli\meter} isotropic de-faced scans of individuals between ages 22 and 35. All scans follow the same MPRAGE protocol with TR \SI{2.4}{\second}, TE \SI{2.14}{\milli\second}, TI \SI{1}{\second}, and flip angle of \ang{8}. The full dataset is available online\footnote{\url{https://www.humanconnectome.org/study/hcp-young-adult}}. The Human Connectome Lifespan Pilot Project (Phase 1a)\footnote{\url{https://www.humanconnectome.org/study-hcp-lifespan-pilot}} is an extension of the Young Adult project and contains imaging data from five age groups. Five participants per age group 25-35, 45-55, and 65-75 were scanned on a 3T Siemens Connectome scanner at an isotropic voxel resolution of \SI{0.8}{\milli\meter} following the Young Adult protocol except for a slightly smaller TE of \SI{2.12}{\milli\second}. In the present study, 30 cases from the Young Adult dataset are used for network training and 20 for validation. A total of 80 cases are used in the final test set. Further, 10 scans from the Lifespan Pilot Project are assembled into a separate test set to assess generalizability to another dataset at \SI{0.8}{\milli\meter} in \Cref{sec:dedicated}.
    
\vspace{1ex}
    \textbf{Rhineland Study}: The Rhineland Study \cite{rhineland} is a large cohort population dataset spanning ages 30 to 90. The \SI{0.8}{\milli\meter} isotropic T1-weighted \ac{MRI} data is acquired on a 3T Siemens Magnetom Prisma scanner using a multi-echo MPRAGE (ME-MPRAGE) sequence with TR \SI{2560}{\milli\second}, 10 TEs (\SI{1.68}{\milli\second}, \SI{3.29}{\milli\second}, \SI{4.90}{\milli\second}, \SI{6.51}{\milli\second}, 6 x \SI{5.0}{\milli\second}), TI~\SI{1100}{\milli\second}, and flip angle \ang{7}. 
    Two separate sets of 30 and 20 subjects are selected for network training and validation, respectively. The final testing set contains another 80 \ac{RS} subjects. 

\vspace{1ex}
All datasets were processed using the open source neuroimage analysis suite FreeSurfer\footnote{\url{http://surfer.nmr.mgh.harvard.edu/}} ~\cite{fischl2002whole,fischl2012freesurfer}. In particular, the FreeSurfer v7.1.1 \ac{HiRes} stream \cite{Zaretskaya2018} was used to generate the desired parcellations following the ``Desikan–Killiany–Tourville'' (DKT) protocol atlas \cite{Klein2012, Desikan2006}. We follow the same label mapping approach as in \FastSurferCNN{} \cite{Henschel_2020}. In short, identical cortical regions on the left and right hemisphere are joined into one class unless they are in close proximity to each other reducing the total number of labels from 95 (DKT without corpus callosum segmentations) to 78 during network training. The affiliation to the left or right hemisphere are restored in the final prediction by estimating the closest \ac{WM} centroid (left or right hemisphere) to each label cluster. A list of all segmentation labels is provided in the Appendix (see \Cref{tab:labels}). In accordance with the FreeSurfer \ac{HiRes} stream, all MRI brain volumes are conformed to a uniform coordinate orientation at their respective isotropic voxel resolutions and robustly normalized to unsigned characters (0..255), i.e.\ the trained network does not depend on skull stripping or bias-field removal. 

\subsection{Voxel size independent neural network}

As mentioned above, inside a standard UNet \cite{Ronneberger2015}, fixed transitions between scales (i.e.\ resolutions) are performed via down- and up-sampling operations (e.g.\ pooling and unpooling, see \Cref{fig:multiresnet}), \new{ naturally leading to a reduction in information}. Theoretically, any scale-transition operation is replaceable with an alternative sampling strategy as long as it still allows gradients to flow effectively through the network. In \mbox{FastSurferVINN}, we use this concept to enforce voxel size independence by changing the first level transition in the encoder and the last in the decoder to a flexible interpolation step (i.e.\ network-integrated resolution-normalization, see \Cref{fig:multiresnet}). Thus, variable transitions between resolutions without restrictions to pre-defined fixed voxel sizes become possible, both during training and inference.

\subsubsection{Network-integrated resolution-normalization}\label{ssec:resolution-normalization}
Similar to spatial transformers \cite{spatialT}, the interpolation-based scale transition can be divided into two parts: (i)~calculation of the sampling coordinates (\textit{grid generator}) and (ii)~interpolation operation (\textit{sampler}) to retrieve the spatially transferred output map. 

The sampling coordinates are calculated based on the scale factor $SF \in \mathbb{R}^+$ -- the quotient of the resolution information of the inner normalized scale $\text{Res}_\text{inner}$ (a tuneable hyperparameter set to \SI{1.0}{\milli\meter} throughout our experiments) and the input image $\text{Res}_\text{native}$. For optimal interpolation, $SF$ is slightly adjusted to ensure integer feature map dimensions. In the first transition step (\Cref{fig:multiresnet}, pre-IDB to IDB), the output feature map $V \in \mathbb{R}^{H_\text{inner}\times W_\text{inner}\times C}$ is produced by sampling the input feature map $U \in \mathbb{R}^{H_\text{native}\times W_\text{native} \times C}$. In the final transition step (\Cref{fig:multiresnet}, CDB to post-CDB), this process is reversed effectively by using the inverse scale factor $1/SF$. 
The interpolation itself is performed by applying a sampling kernel to the input map $U$ to retrieve the value at a particular pixel in the output map $V$. Different interpolation strategies can be defined based on the selection of the sampling kernel. Theoretically, any kernel with definable \mbox{(sub-)}gradients is applicable. Here, we evaluate the bi-linear, bi-cubic, area, and integer sampling kernels (=\ac{NN} interpolation). The sampling is identical for each channel, hence, conserving the spatial consistency. 

\subsubsection{Network architecture modifications}\label{sec:fastsurfer}
The proposed interpolation strategy can, in general, be included in any \ac{CNN} equipped with pooling-based scale transitions. Here, due to its success in neuroanatomical segmentation, the principal network design is based on FastSurferCNN \cite{Henschel_2020} -- a UNet-type network with a series of four competitive dense blocks (CDB) in the encoder and decoder arm separated by a CDB bottleneck layer. In \ac{FastSurferVINN} one additional CDB layer is added (\Cref{fig:multiresnet} to each arm, i.e.\ the pre-IDB and post-CDB).

\textbf{CDB design}
In FastSurferCNN \cite{Henschel_2020}, a CDB is formed by repetitions of the basic composite function consisting of a 5$\times$5 convolution, followed by batch-normalization (BN) and a probabilistic rectified linear unit (pReLU) activation function. In this work we optimize the architecture design slightly by replacing each 5$\times$5 convolution kernel with two 3$\times$3 kernels (see \Cref{fig:multiresnet}). This keeps the effective receptive field size within each block identical to FastSurferCNN while reducing parameter load. We implement this change in \ac{FastSurferVINN} and also, for better comparability, in an updated FastSurferCNN version denoted by: FasturferCNN$^*$. \new{An ablation study detailing the changes from FastSurferCNN to FasturferCNN$^*$ is included in the appendix (\Cref{ssec:appendix_ablation})}.
As in FastSurferCNN, feature competition is achieved by using maxout \cite{maxout} instead of concatenations \cite{densenet} in the local skip connections. In order to guarantee normalized inputs to the maxout activation, the feature map stacking operation is always performed after the BN (see position of maxout in CDB design in \Cref{fig:multiresnet}). 

\textbf{Pre-IDB} 
The additional encoder block in \FastSurferVINN{} (see \Cref{fig:multiresnet}, pre-IDB) transfers image intensity information from the native image to the latent space and encodes voxel size-dependent information before the internal interpolation step. In contrast to the described CDB, the raw inputs are normalized by first passing them through a BN-Conv-BN combination before adhering to the original composite function scheme (Conv-BN-pReLU) (see \Cref{fig:multiresnet}, (pre-)IDB).

\textbf{Post-CDB}
Akin to the pre-IDB, an additional CDB block in the decoder is used to merge the non-interpolated feature information returned from the pre-IDB skip connection and the upsampled feature maps from the network-integrated resolution-normalization step. Both maps are combined via a concatenation operation and then fed to a standard CDB block (see \Cref{fig:multiresnet}, (post-)CDB). After the final 1$\times$1 convolution a softmax operation returns the desired class probabilities.

\subsection{High-res network modifications}
To improve segmentation accuracy of detailed structural features, we explore two network modifications, namely a loss-function weighting scheme and an adaptive attention mechanism.

\textbf{Loss function}

The network is trained with a weighted composite loss function of logistic loss and Dice loss \cite{sdnet}. With $p_{l,i}(x)$ the estimated probability of pixel $i$ to belong to class $l$ and the corresponding ground truth probability $y$, the loss function can be formulated as

\vspace{-2ex}
\begin{equation}
  \mathcal{L} = {\underbrace{-\sum_{l,i} \omega_i y_{l,i} \log p_{l,i}(x)}_\text{Logistic loss}-\underbrace{\sum_{l\vphantom,} \frac{2\sum_i p_{l,i}(x)y_{l,i}}{\sum_i p_{l,i}(x) + \sum_i y_{l,i}}}_\text{Soft Dice loss}}
\label{eq:4a}
\end{equation}
with $\omega_i = \omega_\text{median freq.} + \omega_\text{gradient} + \omega_\text{GM} + \omega_\text{WM/Sulci}$. \begin{figure}[!t]
    \centering
    \includegraphics[width=1.0\columnwidth,keepaspectratio]{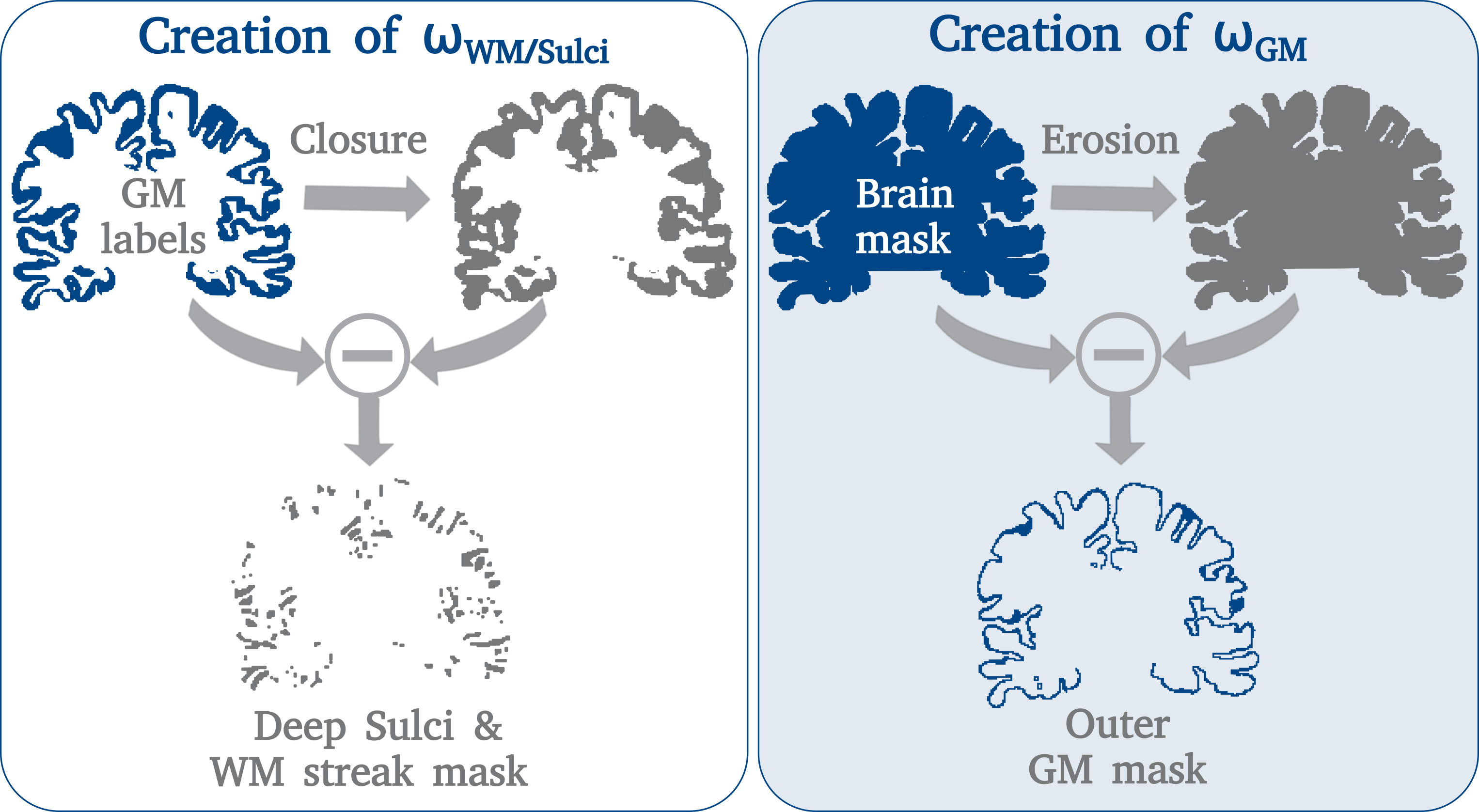}
\caption[Generation of two binary masks for \ac{HiRes} refinement and evaluation]{\ac{HiRes} weight mask generation. Left: The difference between the gray matter label map (blue) and its closure (gray) produces the deep sulci and \ac{WM} strand mask. Right: 
The difference between the original (blue) and eroded (gray) brain mask produces the outer gray matter mask. }
\label{Fig:mask_generation}
\end{figure}

\textbf{Localized weights}
Here, $\omega_\text{median freq.}$ represents median frequency balancing and $\omega_\text{gradient}$ boundary refinement through a 2D gradient vector \cite{sdnet}. We now extend $\omega_i$ by two weighting terms ($\omega_\text{GM}$ and $\omega_\text{WM/Sulci}$) to improve segmentation quality in highly convoluted areas of the cortex that are better represented in submillimeter scans. The \ac{WM} strand and deep sulci mask ($\omega_\text{WM/Sulci}$) emphasizes thin \ac{WM} strands and narrow sulci, and is defined by the voxels added through a binary closing operation on the \ac{GM} labels (left side of \Cref{Fig:mask_generation}). The outer gray matter mask ($\omega_\text{GM}$) accentuates pixels at the boundary of the cortex and is defined by the voxels lost during brain mask erosion (right side of \Cref{Fig:mask_generation}). Overall, $\omega_\text{GM}$ and $\omega_\text{WM/Sulci}$ aim to adjust the underlying decision boundary to closely match the target segmentation in \ac{PVE}-affected locations by assigning higher weights to narrow \ac{WM} strands, deep sulci, and tissue boundaries with emphasis on the border between cortex and \ac{CSF}.
\vspace{1ex}

\textbf{Adaptive attention module}
Context-driven learnable attention mechanisms have been used to automatically select optimal scale or filter sizes for specific image regions and can boost segmentation accuracy of differently sized structures within an image \cite{Qin2018}. As a reference, we therefore evaluate the addition of attention in the pre-encoder and post-decoder. The generation of the activation map follows the method proposed in \cite{Qin2018}. A detailed description is given in the Appendix (\Cref{sec:appendix_attention}). In short, learned activation maps from the attention module are used to dynamically weight each feature response generated by the sequence of convolutions within the CDB. This online weight calculation introduces non-linearities outside the activation function into the CDB. 

\subsection{View aggregation}\label{sec:view_agg}
In order to account for the inherent 3D geometry of the brain, we \new{adopt the same view aggregation scheme as in \cite{Henschel_2020} for all evaluated models}: one F-CNN per anatomical plane is trained and the resulting probability maps are aggregated through a weighted average. Due to missing lateralization in the sagittal view, the number of classes is effectively reduced from 78 to 50 and the weight of the sagittal predictions is reduced by one half compared to the other two views. Inherently, and similar to ensemble learning, the view aggregation combines final soft predictions boosting segmentation accuracy. 

\subsection{Augmentations}\label{sec:scale-aug}
\textbf{External scaling augmentation for \acp{CNN} (exSA)}
The current state-of-the-art method to introduce resolution invariance into neural networks is extensive scale augmentation (see \Cref{fig:networks}B). Therefore, we contrast our proposed network-integrated resolution-normalization against this approach. We use random linear transforms with scaling parameters sampled from a uniform distribution of the predefined range 0.8 to 1.15 to augment images during the training phase. Every minibatch hence consist of a potentially scaled \ac{MRI} (bi-linear interpolation) and a corresponding equally re-scaled label map (\ac{NN} sampling). To disentangle resolution-independence strategies from micro-architecture changes, we use the same CDB and IDB implementations as for \ac{FastSurferVINN} described above, with the exception of the resolution-normalization.
 
\textbf{Internal scaling augmentation for \acp{VINN} (inSA)}
In order to increase the robustness of the latent space interpolation, we augment the scale factor SF introduced by the \ac{VINN} with a parameter $\alpha$, so $SF = \text{Res}_\text{inner}/\text{Res}_\text{native} + \alpha$ during the network-integrated resolution-normalization. This effectively introduces small resolution variations within the grid sampling procedure. The values for $\alpha$ are sampled from a Gaussian distribution with parameters sigma=0.1 and mean=0. Overall, this modification can be \new{interpreted as an internal scale augmentation randomly resizing the feature maps in the latent space} (as opposed to externally augmenting the native images). 

\subsection{Evaluation metrics}
We use the \ac{DSC} and \ac{ASD} to compare different network architectures and modifications against each other, and estimate similarity of the predictions to a number of previously unseen scans with respect to FreeSurfer and manual labels as a reference. Both are standard metrics to evaluate segmentation performance.

Here, the \ac{DSC} \cite{dice_measures_1945,sorensen_method_1948} is defined as twice the intersection of ground truth and prediction divided by the sum of their cardinalities and multiplied by 100. A larger \ac{DSC} represents better overlap between the segmentations with a maximum value of 100 for perfect agreement. The \ac{ASD} measures the average distance (in \si{mm}) between all points $x \in Y, x^{\prime} \in P$ on the outer surface of the ground truth (Y) and the prediction (P).  It is defined as 
\begin{equation}
\begin{aligned}
\text{ASD} = \frac{1}{|Y| + |P|} \left( \sum_{x \in Y} d(x, P) + \sum_{x^{\prime} \in P} d(x^{\prime}, Y) \right)
\label{eq:11}
\end{aligned}
\end{equation} with distance $d(x, P) = \min_{x^{\prime} \in P} \left|\left| x - x^{\prime} \right|\right|_{2}$ representing the minimum of the Euclidean norm. In contrast to the \ac{DSC}, a smaller \ac{ASD} indicates better capture of the segmentation boundaries with a value of zero being the minimum (perfect match). 
Within each section, improvements in segmentation performance are confirmed by statistical testing (Wilcoxon signed-rank test \cite{wilcoxon} after Benjamini-Hochberg correction \cite{Benjamini_Hochberg} for multiple testing) \new{referred to as corrected p throughout the paper}.

\subsection{Training setup}
\textbf{Training dataset: } Due to the nature of the experiments performed in this paper, the dataset composition varies between individual sections. \new{An overview of the different trainingsets} is given in the Appendix (\Cref{tab:datasets}). All representative datasets are balanced with regard to gender and age. Empty slices were filtered from the volumes. \new{All directly compared networks are trained under the same conditions unless stated otherwise. Experimental setups demanding different datasets for training are separated by vertical white lines between bar plots and/or indicated in the figure legend. We additionally discuss this in the corresponding sections}. In the first ablative evaluations of \ac{FastSurferVINN} (\cref{ssec:results-scaling,ssec:results-hires}), 120 representative subjects are selected for training (60 \SI{1.0}{\milli\meter} and 60 submillimeter subjects, see \Cref{tab:datasets}: \textit{Mix}), leaving on average 155 single view planes per subject and a total training size of at least 23k images per network. 
To determine the generalization performance of \ac{FastSurferVINN} across resolutions, the training set is changed such that the submillimeter scans are of the same resolution and study (see \Cref{tab:datasets}: \textit{No \SI{0.7}{\milli\meter}} and \textit{No \SI{0.8}{\milli\meter}}). Similarly, the fixed-resolution networks are trained with 60 or 120 \SI{0.8}{\milli\meter} scans (see \Cref{tab:datasets}: \textit{Only \SI{0.8}{\milli\meter}}). For the \ac{LowRes} version 120 \SI{1.0}{\milli\meter} scans are used (see \Cref{tab:datasets}: \textit{Only \SI{1.0}{\milli\meter}}). In Big-FastSurferVINN, the \SI{1.0}{\milli\meter} component is extended to 1255 scans while keeping the same number (60) of \ac{HiRes} scans as in the original training set (see \Cref{tab:datasets}: \textit{Mix (Big)}).

\textbf{Training parameters: } Independent models for the coronal, axial, and sagittal plane are implemented in \PyTorch~\cite{Paszke2017} and trained for 70 epochs using one NVIDIA V100 GPU with 32 GB RAM. The modified adam optimizer \cite{adamW} is used with a learning rate set to 0.001. A cosine annealing schedule \cite{cosine} adapts the learning rate during training where the number of epochs between two warm restarts is initially set to 10 and subsequently increased by \new{a factor of} two. The momentum parameter is fixed at 0.95 to compensate for the relatively small mini batch size of 16 images. \new{For maximum fairness, all networks presented within this paper have been trained under equal hardware and hyper-parameter settings}.

\section{Results}

We group the presentation of results into two blocks: 1.\ ablative architecture improvements to determine the best performing multi-resolution architecture (\Cref{ssec:results-scaling,ssec:results-hires}), and 2.\ performance analysis to comprehensively characterize the advantages of our \ac{VINN} on a wider variety of datasets, resolutions, scanners, and variations of the training corpus (\Cref{ssec:results-generalizability,sec:dedicated,sec:big}).
Following best practice in data-science, we utilize completely separate datasets during the evaluations: the validation set \Cref{tab:datasets}: Validation (for 1.), and various test sets \Cref{tab:datasets}: Testing (for 2.). This avoids data-leakage and overfitting, i.e., it ensures that training, architectural design decisions, and final testing cannot influence each other, which could lead to overly optimistic results. 

\subsection{Scaling augmentation versus network-integrated resolution-normalization in \ac{FastSurferVINN}}\label{ssec:results-scaling}
The central contribution of this paper is the design and evaluation of a \acl{VINN} for (sub)millimeter whole brain segmentation. Here, we compare segmentation performance of our FastSurferVINN, which avoids interpolation of label maps, with several approaches that rely on traditional scaling data augmentation. Each subsequent improvement in segmentation performance is confirmed by statistical testing (corrected $p < 0.05$).

\begin{figure}[!tb]
    \centering
    \includegraphics[width=\columnwidth,keepaspectratio]{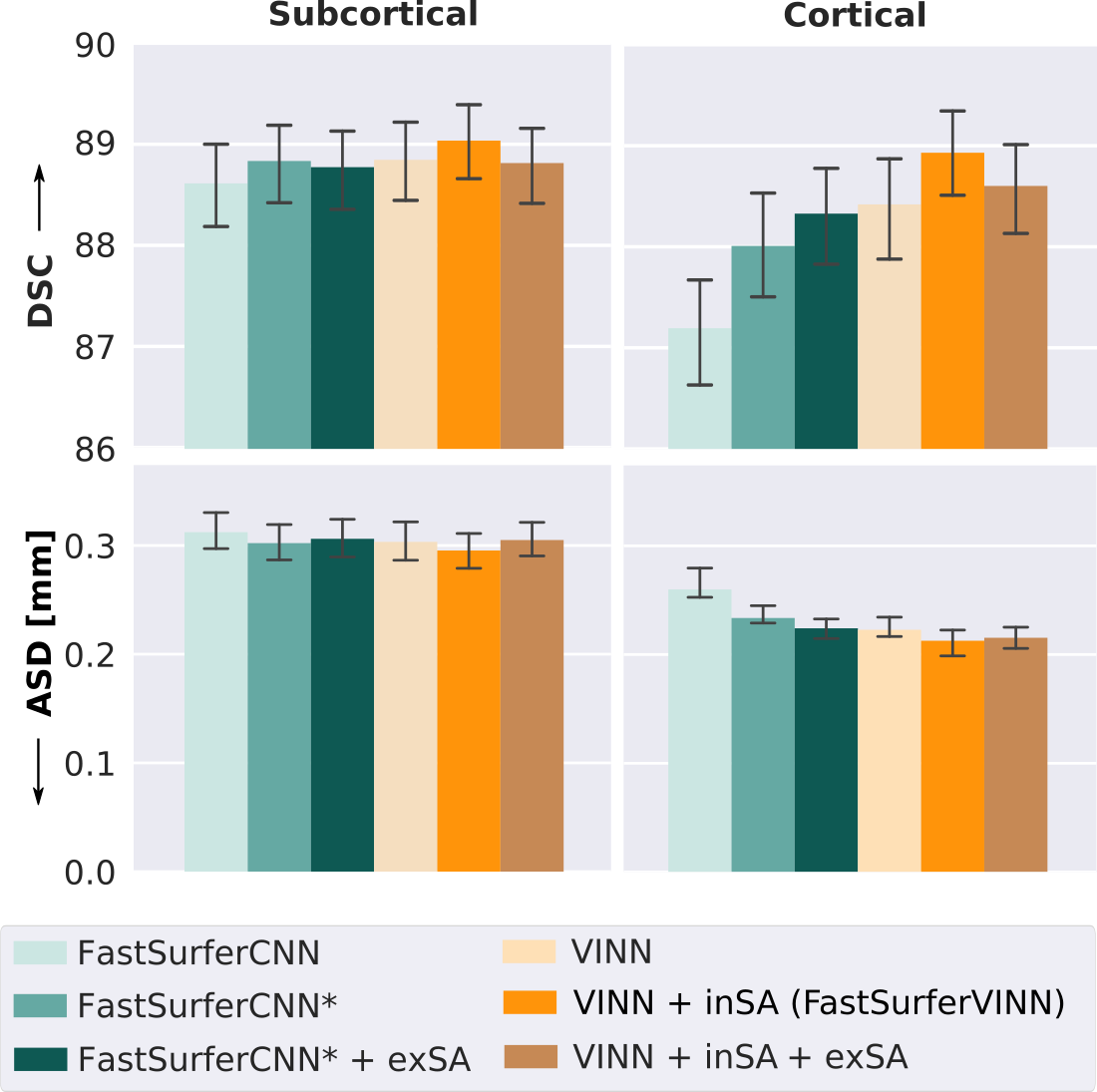}
    \caption{\new{Ablative optimization of FastSurferCNN and comparison to \ac{FastSurferVINN}. FastSurferCNN (light green) is optimized through a switch to 3$\times$3 kernels (\mbox{FastSurferCNN${}^*$}, green). Addition of data augmentation (external scaling augmentation, \mbox{FastSurferCNN${}^*$} + exSA, dark green) improves performance further. \ac{VINN} equipped with internal scaling augmentation (inSA) (\ac{FastSurferVINN}, orange) outperforms all other models on both subcortical (left) and cortical (right) structures with respect to Dice Similarity Coefficient (DSC, top) and average surface distance (ASD, bottom). Further addition of external scaling augmentation negatively affects performance (\ac{VINN} + inSA + exSA).
    Segmentation results with \ac{FastSurferVINN} are significantly better compared to all other models (corrected $p < 10^{-7}$).} }
    \label{fig:s_vs_s}
\end{figure}

Firstly, the original FastSurferCNN CDBs consecutively perform two 5$\times$5 convolution operations followed by a final 1$\times$1 convolution (\Cref{sec:fastsurfer}, FastSurferCNN in \Cref{fig:s_vs_s}). In total, an average \ac{DSC} of 88.63 and a SD of \SI{0.317}{\milli\meter} is reached for the subcortical structures while the cortical structures average around 87.09 for the \ac{DSC} and \SI{0.283}{\milli\meter} for the \ac{ASD}. Optimization of the CDB design (kernel size of 3$\times$3, \Cref{sec:fastsurfer}) leads to a significant increase in the \ac{DSC} and reduction of the \ac{ASD} on both, the subcortical and cortical structures (\Cref{fig:s_vs_s}, \mbox{FastSurferCNN${}^*$}). Particularly, the cortical segmentations are improved with an average \ac{DSC} of 88.01 and a SD of \SI{0.257}{\milli\meter}. Similarly, addition of external scaling augmentation (exSA) to \mbox{FastSurferCNN${}^*$} significantly improves segmentation accuracy on the cortical structures (\Cref{fig:s_vs_s}, \mbox{FastSurferCNN${}^*$} + exSA). Performance on the subcortical structures is, however, slightly reduced. 

Interestingly, \ac{VINN}, which avoids label map interpolation altogether, shows a positive effect across structures compared to \mbox{FastSurferCNN${}^*$} with and without scaling augmentation (\Cref{fig:s_vs_s}, \ac{VINN}). Specifically, combination of \ac{VINN} with our internal scaling augmentation (\ac{VINN} + inSA, referred to as \ac{FastSurferVINN}) strengthens segmentation performance significantly. The \ac{DSC} increases to 89.05 for the subcortical and 88.93 for the cortical structures representing the best value across the compared architectures. Similarly, the \ac{ASD} is significantly reduced with a distance of \SI{0.293}{\milli\meter} for the subcortical and \SI{0.226}{\milli\meter} for the cortical structures (\Cref{fig:s_vs_s}, \ac{FastSurferVINN}). Finally, addition of external scaling augmentation negatively impacts segmentation performance of \ac{VINN} (\Cref{fig:s_vs_s}, \ac{VINN} + inSA + exSA). On average, performance on the subcortical structures is reduced to the level of \mbox{FastSurferCNN${}^*$} + exSA, while the cortical structures still represent the second best result for both, \ac{DSC} and \ac{ASD}. \new{Overall, \ac{FastSurferVINN} outperforms all traditional scale augmentation approaches by a significant margin (corrected $p<10^{-7}$).} 

\begin{figure}[!bt]
    \centering
    \includegraphics[width=\columnwidth,keepaspectratio]{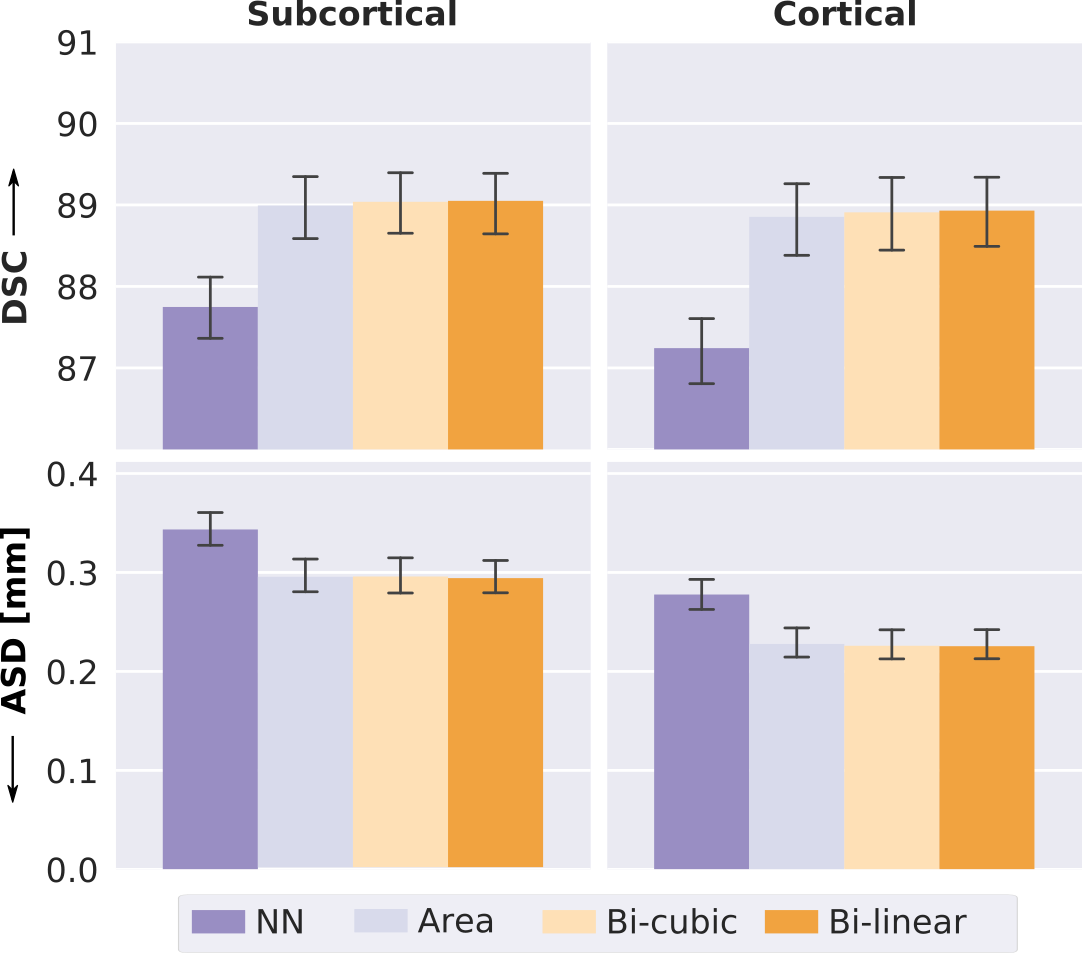}
    \caption{Effect of sampling kernels on network-integrated resolution-normalization. Comparison of nearest-neighbour (NN, purple), area (light violet), bi-cubic (light yellow), and bi-linear (orange) sampling kernels with respect to the Dice Similarity Coefficient (DSC, top) and the average surface distance (ASD, bottom) for subcortical (left) and cortical (right) structures. 
    \new{Segmentation performance with NN is significantly worse than all other interpolation strategies (corrected $p < 10^{-13}$). Area, bi-cubic and bi-linear give equivalent results.}}
    \label{fig:interpol}
\end{figure}

\new{As our network-integrated resolution-normalization directly operates on continuous 2D feature maps in the latent space}, various sampling kernels can be incorporated. Here, we evaluate the effect of four different interpolation strategies, namely \ac{NN}, area, bi-cubic, and bi-linear.
As visible in \Cref{fig:interpol}, with the exception of \ac{NN} changing the sampling kernel does not significantly affect segmentation performance. The \ac{NN} interpolation (\Cref{fig:interpol}, first column) overall reduces performance by ~2\% for \ac{DSC}. The change on the \ac{ASD} is more severe with a decrease in performance by 13\% for the subcortical and 7\% for the cortical structures. The bi-linear kernel (\Cref{fig:interpol}, last column) shows the best performance overall with a 89.05 \ac{DSC} and an \ac{ASD} of \SI{0.293}{\milli\meter} for the subcortical structures and a 88.93 \ac{DSC} and an \ac{ASD} of \SI{0.226}{\milli\meter} for the cortical structures. Given that the bi-linear interpolation is computationally favourable to bi-cubic, we henceforth keep it as the sampling kernel of choice. 

\subsection{\ac{HiRes} specific adjustments}\label{ssec:results-hires}

The higher resolution in submillimeter scans reduces \acp{PVE} and offers a potential to optimize segmentation performance (see \Cref{fig:PVE}). Overall, the changes in tissue and border assignment might, however, not be accurately captured during network training as they represent marginal changes in the loss compared to the whole brain volume. In order to focus on the \ac{PVE} affected regions (specifically small \ac{WM} strands and deep sulci), we test two different modifications. First, a \ac{HiRes} loss function (HiRes Loss) up-weights information from the areas in question. Second, attention mechanisms are introduced to enable automatic refocusing on important information during network training. Here, we equip \ac{FastSurferVINN} separately with both modifications and evaluate the change in segmentation performance with respect to the \ac{DSC} and \ac{ASD}.

\begin{figure}[!t]
    \centering
    \includegraphics[width=\columnwidth,keepaspectratio]{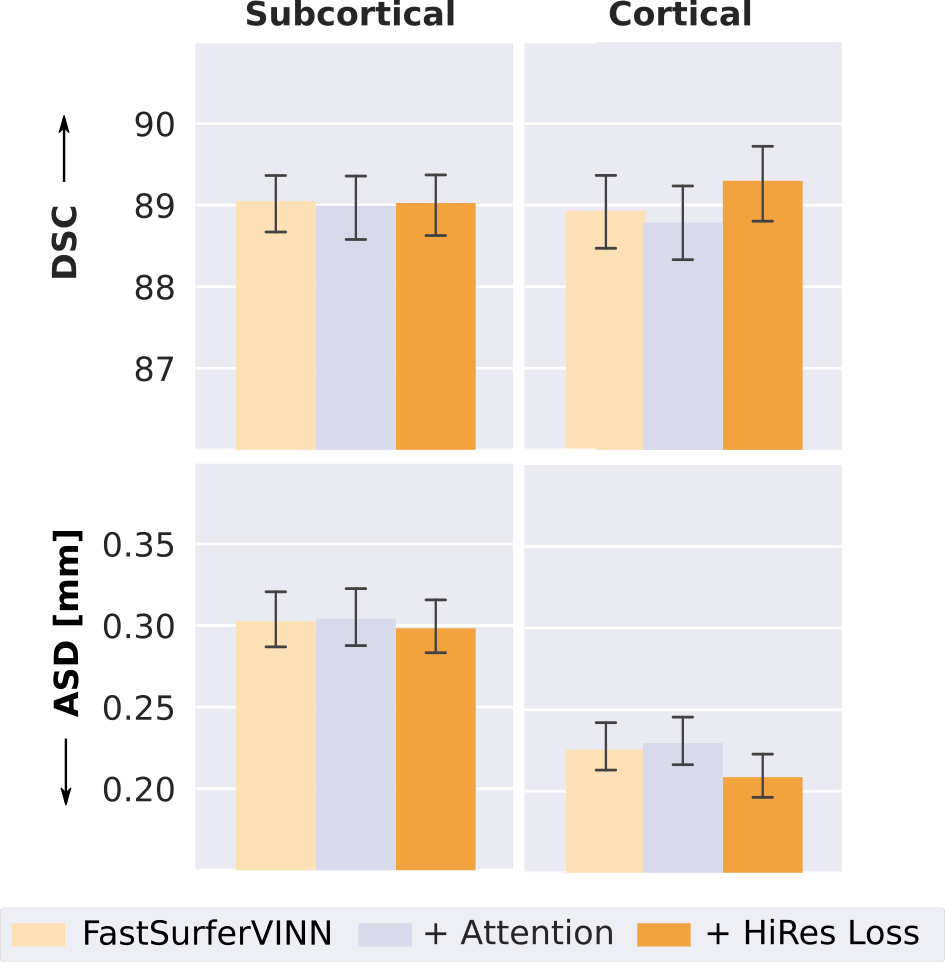}
    \caption{\new{Adaptation of the original loss function through addition of \ac{HiRes} weights focusing on areas strongly effected by \acp{PVE} (HiRes Loss, right bar) significantly improves segmentation performance on the cortical structures (compared to \ac{FastSurferVINN} with original loss, left bar). Addition of attention (middle bar) does not lead to a significant improvement compared to the baseline. Dice Similarity Coefficient (DSC, top) and average surface distance (ASD, bottom) are shown for subcortical (left) and cortical (right) structures.
    Cortical structures are significantly better segmented with the HiRes Loss (corrected $p < 10^{-5}$). No significant change was detected on the subcortical structures.}
    }
    \label{fig:hires}
\end{figure}

Training \ac{FastSurferVINN} with the new \ac{HiRes} loss function (see \Cref{fig:hires}, + \ac{HiRes} Loss, right bar) significantly improves segmentation performance on the cortical structures (corrected $p < 10^{-5}$), while maintaining high accuracy on the subcortical structures (no significant change consistent with expectations). A final \ac{DSC} of 89.3 and an \ac{ASD} of \SI{0.209}{\milli\meter} is achieved for the cortical structures. Further addition of attention, while simultaneously adjusting the number of feature maps to keep the total number of network parameters of \ac{FastSurferVINN} constant, does not lead to a significant change in segmentation performance (see \Cref{fig:hires}, + attention, middle bar). Due to the overall significant improvement on \ac{DSC} and \ac{ASD}, the HiRes Loss modification is included in all further comparisons.

\begin{figure*}[!t]
    \centering
    \includegraphics[width=\textwidth,keepaspectratio]{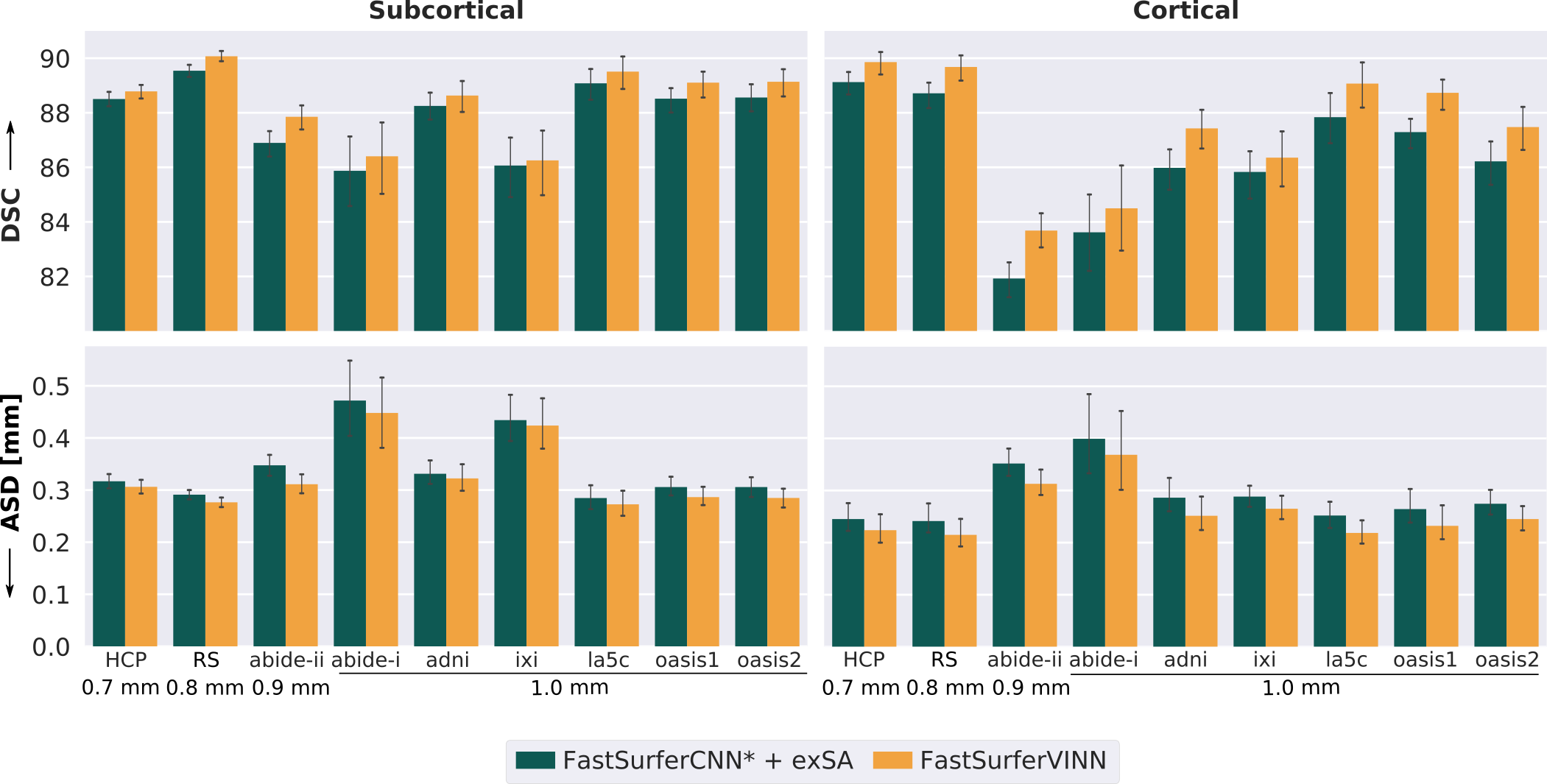}
    \caption{Improved generalization performance of \ac{FastSurferVINN} across nine datasets. \ac{FastSurferVINN} (orange) outperforms \mbox{FastSurferCNN${}^*$} + external scale augmentation (+exSA, dark green) across subcortical (left) and cortical structures (right) with respect to to Dice Similarity Coefficient (DSC, top) and average surface distance (ASD, bottom). Results are consistently better for all datasets (\ac{HCP}, \ac{RS}, \ac{abide-ii}, ABIDE-I, ADNI, IXI, LA5C, OASIS1 and OASIS2). }
    \label{fig:gen}
\end{figure*} 

\subsection{Generalizability}\label{ssec:results-generalizability}
After selection of the best architectures for both, VINN and CNN, we now perform a detailed evaluation in a broad selection of test scenarios (see \Cref{tab:datasets}: Testing). \new{To highlight the advantages of our VINN, we compare \ac{FastSurferVINN} with \mbox{FastSurferCNN${}^*$} + exSA. The latter is improved over the state-of-the-art FastSurferCNN by architectural updates. The difference between FastSurferCNN${}^*$ + exSA and FastSurferVINN is exclusively the substitution of external scale augmentation (exSA, \cref{sec:scale-aug}) with network-integrated resolution-normalization (see \cref{ssec:resolution-normalization}) and internal scale augmentation (inSA, \cref{sec:scale-aug})}.

\subsubsection{Across datasets}\label{sec:generalization}
In this section, we evaluate the generalization capabilities of \ac{FastSurferVINN} in comparison to scale augmentation on the test corpus to establish the performance metrics across multiple datasets. 

\ac{FastSurferVINN} consistently reaches the best \ac{DSC} and \ac{ASD} across all nine datasets (\Cref{fig:gen}, orange bar). All improvements compared to external scale augmentation (dark green bar) are again significant (corrected $p < 0.01$). The best performance is reached for the submillimeter scans from \ac{HCP} and \ac{RS} with a subcortical \ac{DSC} of 88.78 and 90.07 and a cortical \ac{DSC} of 89.86 and 89.68, respectively. La5c reaches the highest \ac{DSC} for the \SI{1.0}{\milli\meter} scans (89.51 and 89.07 subcortical and cortical \ac{DSC}). Similar to the \ac{DSC}, the \ac{ASD} is significantly improved for \ac{FastSurferVINN} compared to the scale augmentation by around 5.2\% for the subcortical and 10.5\% for the cortical structures.  On the cortical and subcortical structures the best \ac{ASD} is reached for \ac{HCP} (\SI{0.31}{\milli\meter}, \SI{0.22}{\milli\meter}), \ac{RS} (\SI{0.28}{\milli\meter}, \SI{0.21}{\milli\meter}) and la5c (\SI{0.27}{\milli\meter}, \SI{0.22}{\milli\meter}). The biggest improvement with regards to comparability between \ac{FastSurferVINN} and scale augmentation can be seen for \ac{abide-ii}. Here, the \ac{DSC} differs by 1.1 and 2.1\% for the subcortical and cortical \ac{DSC} and around 11\% for the \ac{ASD}. This again reflects the better cross-resolution generalization performance of \ac{FastSurferVINN} already outlined in the previous section. ADNI and OASIS1 also benefit strongly from \ac{FastSurferVINN}, specifically for the cortical structures. Here, the \ac{DSC} and \ac{ASD} are improved by around 1.66\% and 12.1\%, respectively. Overall, the cortical structures benefit stronger from the internal resolution-normalization in \ac{FastSurferVINN}.

\begin{figure}[!t]
    \centering
    \includegraphics[width=\columnwidth,keepaspectratio]{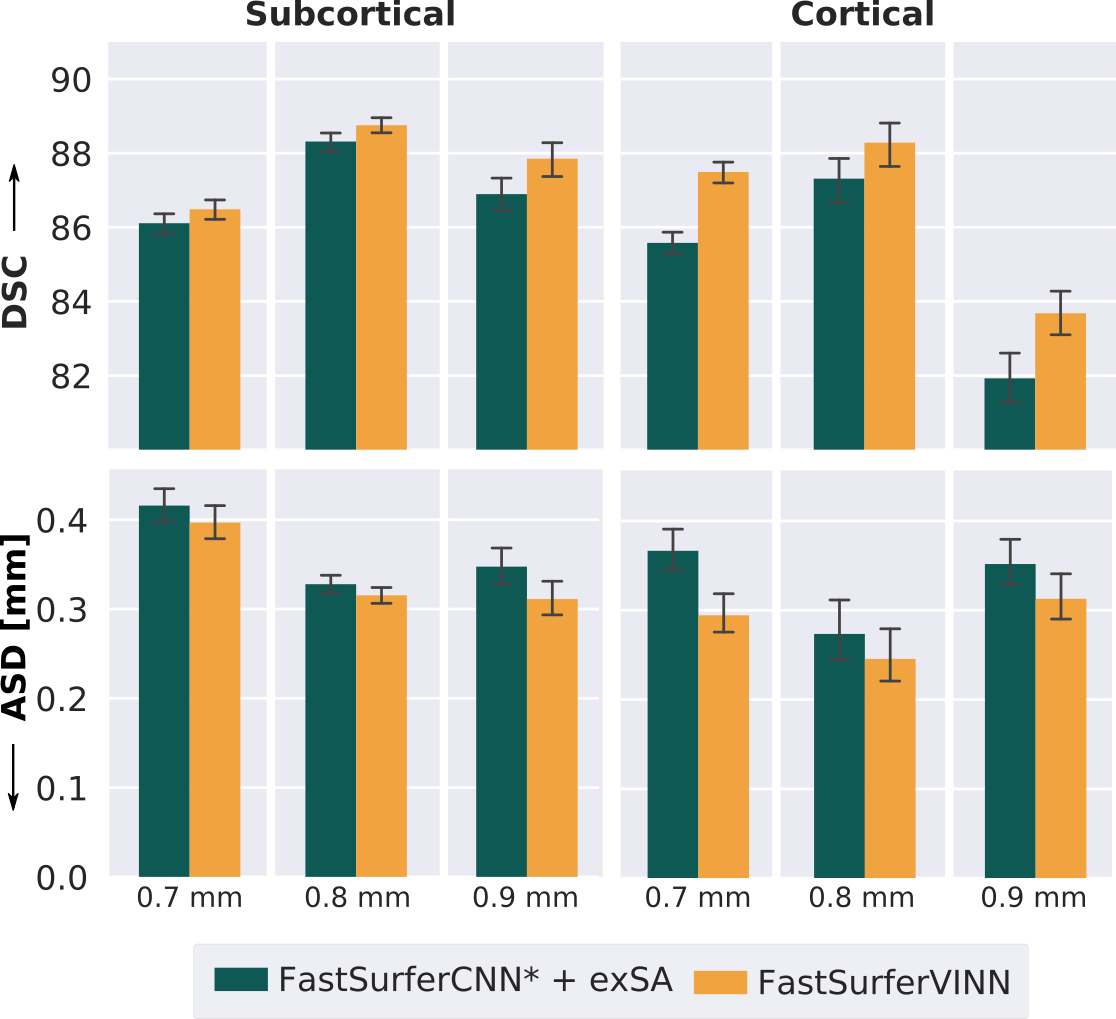}
    \caption{Improved generalization performance of \ac{FastSurferVINN} to resolutions not encountered during network training \new{(here training datasets are customized, see \Cref{sec:loo,sec:appendix_datasets})}. \ac{FastSurferVINN} (orange) outperforms \mbox{FastSurferCNN${}^*$} equipped with external scale augmentation (+ exSA, dark green) with respect to Dice Similarity Coefficient (DSC, top) and average surface distance (ASD, bottom) across subcortical (left) and cortical structures (right). Results are significantly better across all resolutions (\SI{0.7}{\milli\meter}, \SI{0.8}{\milli\meter}, and \SI{0.9}{\milli\meter}, corrected $p < 10^{-4}$).}
    \label{fig:resolution}
\end{figure}

\subsubsection{Unseen resolutions}\label{sec:loo}
A core aspect of \ac{FastSurferVINN} is the implicit compatibility with a variety of resolutions independent of their explicit presence in the training corpus. In order to investigate the generalization capacity of \ac{FastSurferVINN} in contrast to scale augmentation approaches, we evaluate the inter- and extrapolation capabilities of the trained networks based on the segmentation performance for three resolutions purposefully excluded during network training.

To this end, we specifically drop either (i)~all 30 \SI{0.8}{\milli\meter}, or (ii)~all 30 \SI{0.7}{\milli\meter} scans from the training corpus (see \Cref{tab:datasets}: \textit{No \SI{0.7}{\milli\meter}} and \textit{No \SI{0.8}{\milli\meter}}). In order to ensure comparability with respect to the total number of data points and balance between \ac{HiRes} and \ac{LowRes} scans, an equal number of subjects from the other respective \ac{HiRes} datasets are added. In addition, we evaluate performance using the original mixed training set on 25 subjects from \ac{abide-ii}, representing an unseen scanner type (Philips) and resolution (\SI{0.9}{\milli\meter}).

As presented in \Cref{fig:resolution}, \ac{FastSurferVINN} (orange bar) consistently outperforms traditional scale augmentation (dark green bar) across all resolutions (corrected $p < 10^{-4}$). Segmentation performance on the \SI{0.7}{\milli\meter} scans reflects the network's resolution extrapolation capabilities (training corpus consists of \SI{0.8}{\milli\meter} and \SI{1.0}{\milli\meter} scans only). Here, \ac{FastSurferVINN} reaches a \ac{DSC} of 86.49 and 87.50 for the subcortical and cortical structures, respectively, representing a significant increase compared to the traditional scale augmentation approach. The improvements on the \ac{ASD} are even more pronounced with \ac{FastSurferVINN} reducing the \ac{ASD} by 4.5\% to \SI{0.397}{\milli\meter} on the subcortical and by 19.7\% to \SI{0.294}{\milli\meter} on the cortical structures. Comparison of the interpolation capabilities reflected in the \SI{0.8}{\milli\meter} results (training corpus consists of \SI{0.7}{\milli\meter} and \SI{1.0}{\milli\meter}) paint a similar picture. \ac{FastSurferVINN} reaches the highest \ac{DSC} and lowest \ac{ASD} for both subcortical (\ac{DSC}: 88.75, \ac{ASD}: \SI{0.316}{\milli\meter}) and cortical structures (\ac{DSC} of 88.28 and \ac{ASD} of \SI{0.273}{\milli\meter}). The difference to traditional scale augmentation approaches is again more evident on the cortical structures. Finally, metrics on the \SI{0.9}{\milli\meter} Philips scans are significantly better with \ac{FastSurferVINN}. A final \ac{DSC} of 87.85 and 83.68 and an \ac{ASD} of \SI{0.311}{\milli\meter} and \SI{0.313}{\milli\meter} is reached on the subcortical and cortical structures, respectively. 

\new{We face the limitation that no publicly available 3T datasets exists at finer resolutions than \SI{0.7}{\milli\meter}. In order to further explore the extrapolation capabilities of FastSurferVINN in comparison to traditional scale augmentation (\mbox{FastSurferCNN${}^*$} + exSA), we therefore evaluate segmentation performance at lower resolutions vastly outside the training range (\SI{1.4}{\milli\meter} and \SI{1.6}{\milli\meter}). Note, that FreeSurfer segmentations are not available at lower native resolutions either, as images at resolutions coarser than \SI{1.0}{\milli\meter} are upsampled to a millimeter voxel resolution in an initial conversion step \cite{fischl2012freesurfer}. Therefore, we downsample the high-resolution FreeSurfer segmentations of \ac{HiRes} test sets (\ac{HCP} and \ac{RS}) by a factor of 2 along each axis using majority voting to generate ``ground truth''. The intensity images used for inference were resampled using cubic interpolation. }

\begin{figure}[!b]
    \centering
    \newfigure{\includegraphics[width=0.7\columnwidth,keepaspectratio]{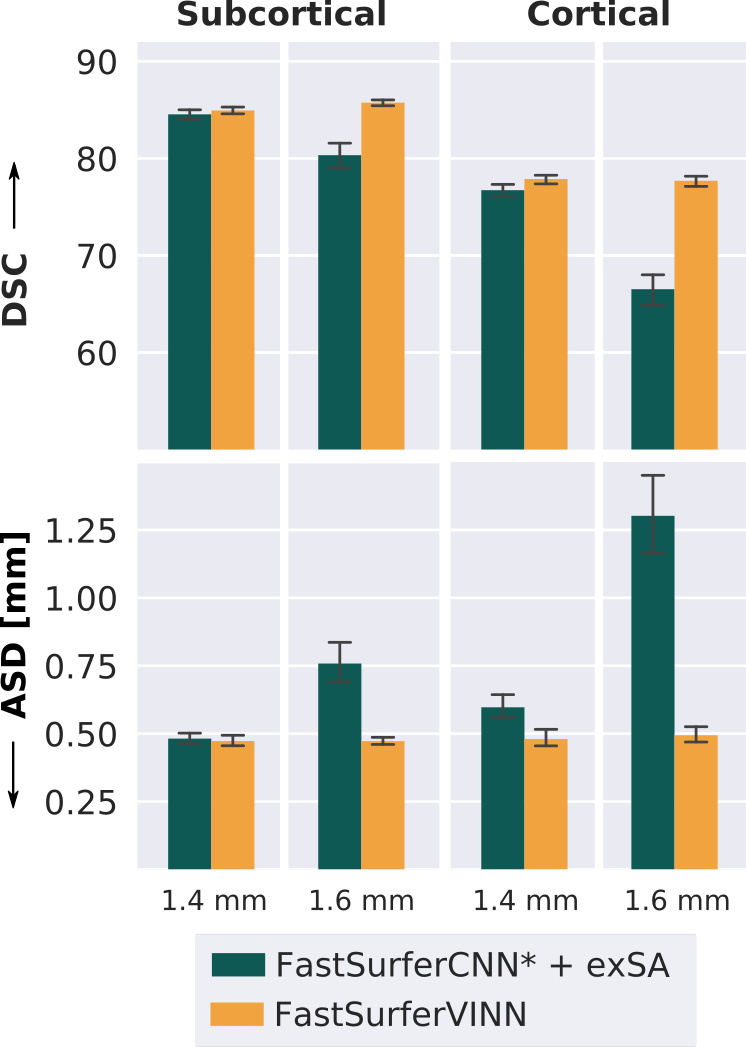}}
    \caption{\new{Superior generalization performance of \ac{FastSurferVINN} to resolutions vastly outside the training domain (\SI{1.4}{\milli\meter}, \SI{1.6}{\milli\meter}). \ac{FastSurferVINN} (orange) outperforms scale-augmentation (\mbox{FastSurferCNN${}^*$} + exSA, green) highlighting its extrapolation capabilities. Results are significantly better with respect to Dice Similarity Coefficient (DSC, top) and average surface distance (ASD, bottom) across subcortical (left) and cortical structures (right) (corrected $p < 0.001$ for 1.4 mm and $p < 10^{-13}$ for 1.6 mm).}}
    \label{fig:resolution_lr}
\end{figure} 

\new{In \Cref{fig:resolution_lr} we illustrate a strong divergence between FastSurferVINN (orange) and FastSurferCNN* + exSA (dark green) as we test with resolutions increasingly outside the training range. With FastSurferVINN, accuracy stays consistently high for both, \SI{1.4}{\milli\meter} and \SI{1.6}{\milli\meter} downsampled images. Here, FastSurferVINN reaches a \ac{DSC} of above 84.90 and a \ac{ASD} of \SI{0.473}{\milli\meter} on the subcortical structures (left plot) and a \ac{DSC} of around 77.70 (77.88 on the \SI{1.4}{\milli\meter} and 77.69 on the \SI{1.6}{\milli\meter} images) and an \ac{ASD} of \SI{0.480}{\milli\meter} and \SI{0.494}{\milli\meter} on the cortical structures (right plot). The difference to FastSurferCNN* + exSA is again more evident on the cortical structures. Overall, FastSurferVINN outperforms the scaling augmentation by 6.74 \% on the subcortical and 16.80 \% on the cortical structures with respect to the \ac{DSC} when using the downsampled \SI{1.6}{\milli\meter} image as an input. Note, that this comparison is solemnly performed to visualize the performance difference between VINN and augmentation in the absence of high-resolution ground truth beyond the training range. We expect a reduction of performance at low resolutions due to stronger \ac{PVE} and an 8-fold information reduction.}

\begin{figure}[!bt]
    \centering
    \includegraphics[width=\columnwidth,keepaspectratio]{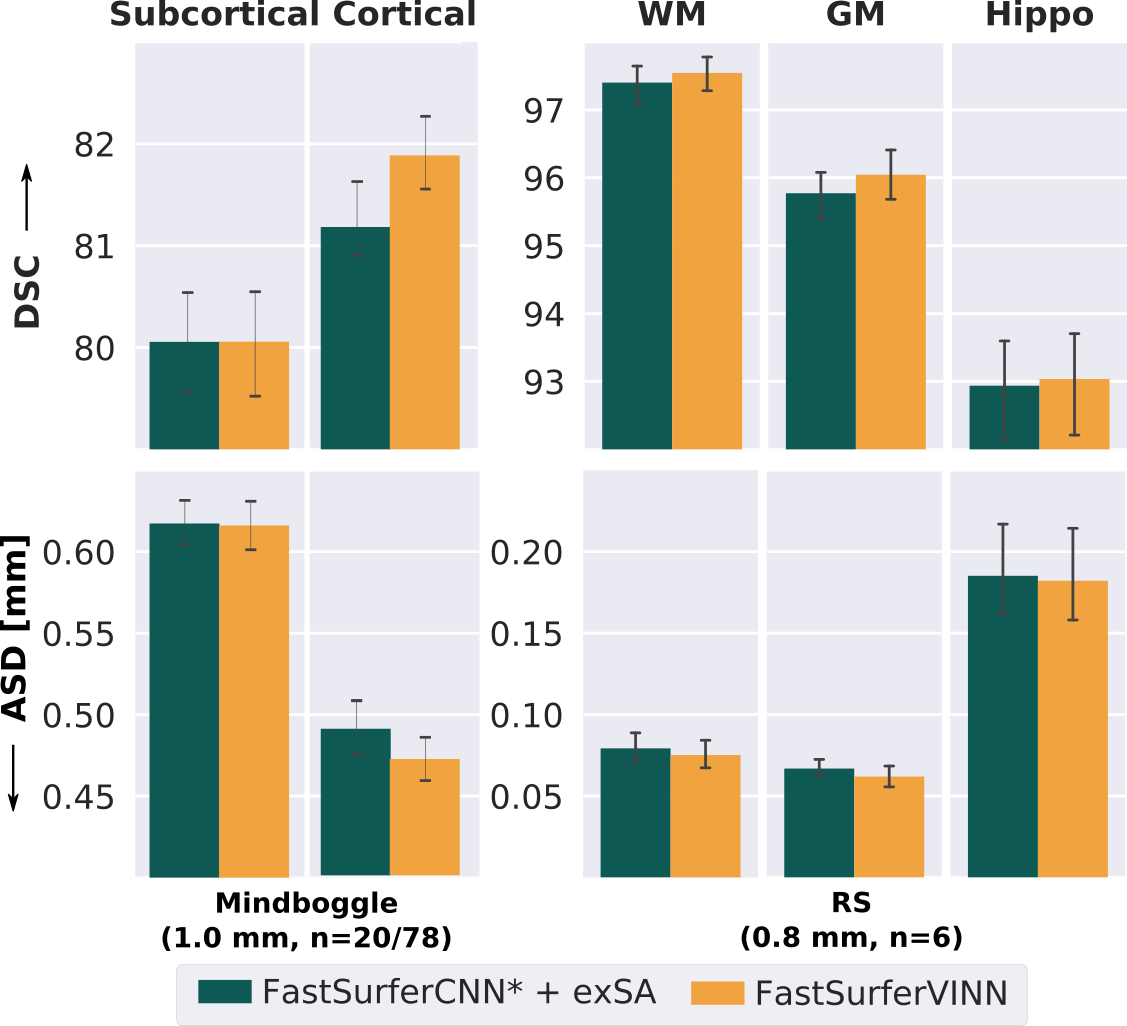}
    \caption{Performance of \ac{FastSurferVINN} with respect to manual references. Based on the \SI{1.0}{\milli\meter} scans in Mindboggle\-101 (left plot) \ac{FastSurferVINN} (orange) outperforms external scale augmentation (+exSA, dark green) on the cortical structures (right, $N=78$) with respect to Dice Similarity Coefficient (DSC, top) and average surface distance (ASD, bottom). Results on the subcortical structures (left side, $N=20$) are equivalent for both approaches. Similarly, segmentation results are better for the \SI{0.8}{\milli\meter} scans of the \ac{RS} (right plot, $N=6$) for white matter (WM), gray matter (GM), and hippocampus (Hippo). }
    \label{fig:mindboggle_eval}
\end{figure}
\subsubsection{Comparison to manual reference}\label{sec:manual}
Due to the limited availability of manual labels, FreeSurfer segmentations have been used as the reference for comparison so far. In order to account for potential biases we now also evaluate \ac{DSC} and \ac{ASD} for the manually edited cortical regions on 78 subjects from the \SI{1.0}{\milli\meter} Mindboggle\-101 dataset \cite{Klein2012}. The subcortical segmentations available for a subset of 20 subjects within this cohort are used for subcortical evaluations. 

On the subcortical structures (see \Cref{fig:mindboggle_eval}, left part of left plot), \ac{FastSurferVINN} (orange bar) and traditional scale augmentation (dark green bar) perform equally well with respect to both, \ac{DSC} (80.26) and \ac{ASD} (\SI{0.616}{\milli\meter}). The difference between the two methods is more pronounced on the cortical structures (\Cref{fig:mindboggle_eval}, right side of left plot). \ac{FastSurferVINN} reaches a final \ac{DSC} of 81.89 and an \ac{ASD} of \SI{0.471}{\milli\meter}), representing a significant improvement compared to the external scale augmentation approach (corrected $p < 0.005$). These results again indicate that the cortical structures specifically benefit from the internal interpolation approach of \ac{FastSurferVINN}. 

\begin{figure*}[!hbt]
    \centering
    \newfigure{\includegraphics[width=\textwidth,keepaspectratio]{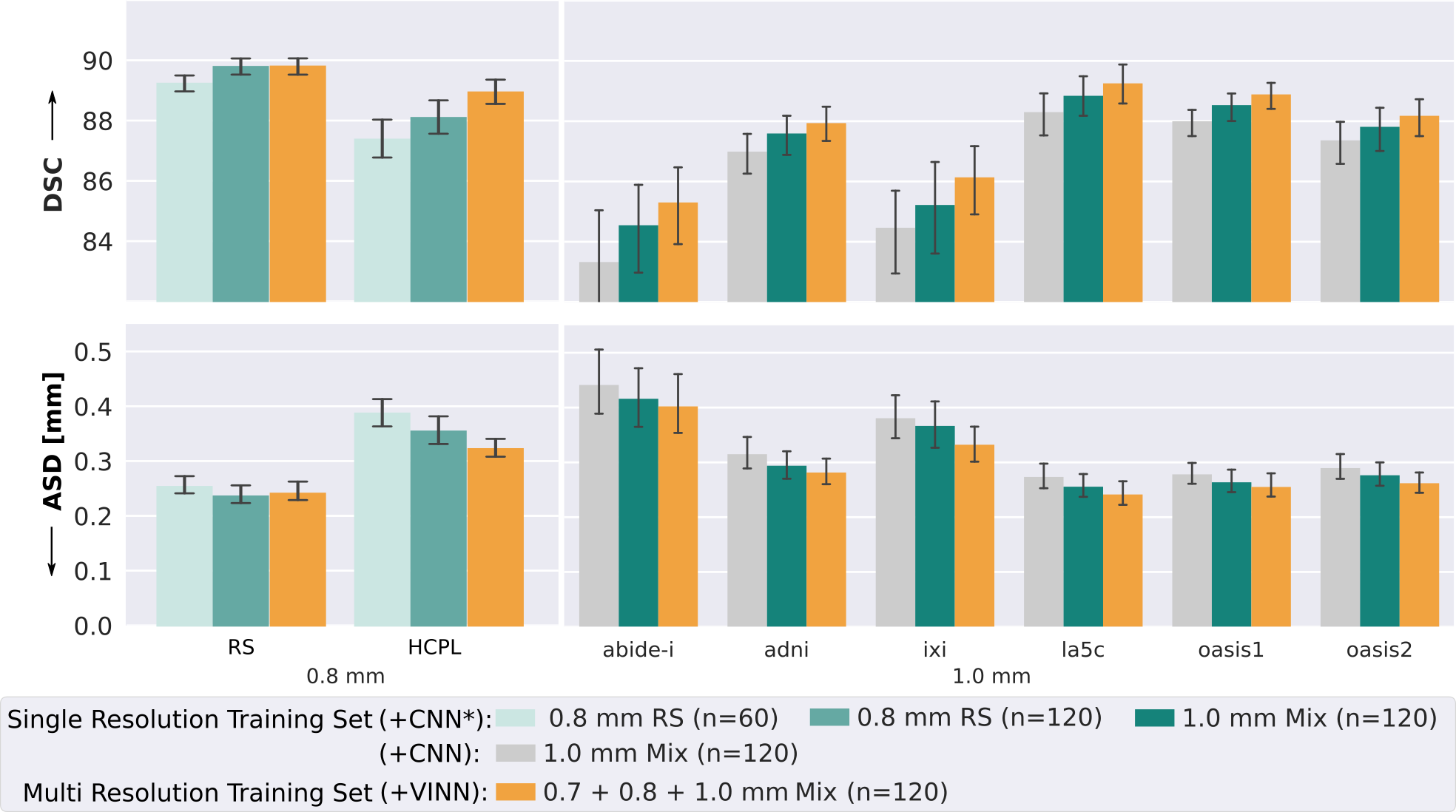}}
    \caption{Flexible- versus fixed-resolution networks. \ac{FastSurferVINN} (orange) is comparable to, or outperforms, all fixed-resolution networks (green) (left plot: \SI{0.8}{\milli\meter} from \ac{RS} only, right plots: \SI{1.0}{\milli\meter}) with respect to the Dice Similarity Coefficient (DSC, top) and average surface distance (ASD, bottom). On the submillimeter scans (left plots) generalization to an unseen dataset (HCPL) is significantly improved. Results are consistently better for the \SI{1.0}{\milli\meter} scans (right plot). \new{To highlight cumulative VINN and architectural optimizations, we also compare with the state-of-the-art FastSurferCNN (gray, without optimizations from \cref{ssec:results-scaling,ssec:results-hires}, which are already included in CNN${}^*$). We retrain this \SI{1}{\milli\meter} fixed-resolution network ensuring equal training datasets and conditions.}}
    \label{fig:dedicated}
\end{figure*}

In order to evaluate the benefit of \ac{FastSurferVINN} for submillimeter scans, we compare segmentation accuracy on three structures (\ac{WM}, \ac{GM} and hippocampus) on an in-house set, as no manual \ac{HiRes} full brain segmentations are publicly available. \new{Note, while we previously reported averages across the 45 cortical and 33 subcortical structures listed in \cref{tab:labels} for clarity, we now calculate performance measures specifically for these three individual structures. To generate manual annotations, a trained expert corrected FreeSurfer generated segmentations }on six cases from the \ac{RS}. The results are depicted in the left part of \Cref{fig:mindboggle_eval}. \ac{FastSurferVINN} again outperforms traditional scale augmentation with a final \ac{DSC} of 97.54, 96.04, and 93.04 and \ac{ASD} of \SI{0.075}{\milli\meter}, \SI{0.062}{\milli\meter}, \SI{0.181}{\milli\meter} for \ac{WM}, \ac{GM}, and hippocampus, respectively. Consistent with previous results, the \ac{GM} segmentations show the strongest improvement. Due to the small sample size, no statistical analysis could be performed.

\subsection{Fixed-resolution networks versus \ac{FastSurferVINN}}\label{sec:dedicated}

In order to analyse the advantage of the multi-resolution training approach over a fixed-resolution network, we compare \ac{FastSurferVINN} with networks trained on (i)~60 \SI{0.8}{\milli\meter} subjects, (ii)~120 \SI{0.8}{\milli\meter} subjects, and (iii)~120 \SI{1.0}{\milli\meter} subjects (see \Cref{tab:datasets}: \textit{Only \SI{0.8}{\milli\meter}} and \textit{Only \SI{1.0}{\milli\meter}}). Inherently, this analysis highlights how the interplay of \ac{HiRes} information from submillimeter 3T scans and dataset-variations from \ac{LowRes} \acp{MRI} can mutually benefit segmentation performance on the different resolutions. 
Overall, the analysis of \Cref{fig:dedicated} highlights three important properties. 

First, increasing the number of data samples has a strong effect on segmentation performance. The fixed-resolution network trained with 60 \SI{0.8}{\milli\meter} scans (left plot; light green bar) reaches an average \ac{DSC} of 89.28 and \ac{ASD} of \SI{0.257}{\milli\meter} for \ac{RS} data (same cohort as in the training set). Doubling the number of cases to 120, significantly improves both measures (corrected $p<10^{-9}$) with a final \ac{DSC} of 89.85 and \ac{ASD} of \SI{0.245}{\milli\meter} (\ac{FastSurferVINN}, orange bar). Interestingly, \ac{FastSurferVINN} trained with 60 \ac{HiRes} (30 \SI{0.7}{\milli\meter} and 30 \SI{0.8}{\milli\meter}) and 60 \ac{LowRes} scans (\SI{1.0}{\milli\meter}) reaches the same accuracy, as if these scans were actually all \ac{HiRes} scans from the \ac{RS} cohort (120 \SI{0.8}{\milli\meter}, dark green bar). Benefits from the sample size increase are, therefore, independent of the data resolution. Advantageously, the additional 90 \ac{HiRes} cases at \SI{0.8}{\milli\meter} may easily be integrated into the \ac{FastSurferVINN} training dataset, hence, increasing the training corpus with another expected performance gain. 

Second, fixed-resolution networks perform well on the cohort they are trained on, but lack generalization capability to other datasets (left plot, HCPL). When comparing networks trained with equally sized datasets (120 \ac{HiRes} \ac{RS} fixed-resolution net (dark green bar) and \ac{FastSurferVINN} (orange bar)) on the \SI{0.8}{\milli\meter} \ac{HCP}L cohort, \ac{FastSurferVINN} clearly outperforms the fixed-resolution network with an increase in \ac{DSC} to 88.99 and \new{a decrease of the} \ac{ASD} to \SI{0.326}{\milli\meter}, representing a significant improvement (corrected $p <0.005$). The gap in segmentation accuracy between the \ac{RS} and \ac{HCP} cohort with respect to \ac{DSC} is halved for \ac{FastSurferVINN} compared to the fixed-resolution approach and a similar reduction is visible for the difference in \ac{ASD}. 
Third, multi-resolution training benefits \SI{1.0}{\milli\meter} datasets as well. \new{In the right part of \Cref{fig:dedicated}, we compare \ac{FastSurferVINN} (orange bar) and two fixed-resolution \SI{1.0}{\milli\meter} networks -- the original FastSurferCNN (CNN, gray bar) as well as the optimized FastSurferCNN* (CNN*, dark green bar) -- with \ac{DSC} (top) and \ac{ASD} (bottom) metrics}. \new{FastSurferCNN is included to highlight the cumulative performance gain of FastSurferVINN (architectural optimization already included in CNN* and voxel-size independence via resolution normalization)}. The training sets of \new{all three} networks contain the exact same subjects, with the only difference being the \acp{MRI} resolution: 
We train the fixed-resolution networks exclusively with \SI{1.0}{\milli\meter} images (native or downsampled from \ac{HiRes}), while \ac{FastSurferVINN} uses all images at their native resolution (\SI{1.0}{\milli\meter}, \SI{0.8}{\milli\meter}, or \SI{0.7}{\milli\meter}). Note, since the FreeSurfer-based label maps are obtainable at either resolution, we circumvent resampling with \ac{NN}. Interestingly, the \ac{HiRes} information from submillimeter scans boosts performance for the \SI{1.0}{\milli\meter} scans. \ac{FastSurferVINN} reaches a \ac{DSC} of 87.62 and an \ac{ASD} of \SI{0.296}{\milli\meter} representing a significant improvement in \ac{DSC} and \ac{ASD} compared to the fixed-resolution networks (corrected $p<0.001$). \new{Compared to the original FastSurferCNN, on average an improvement of 1.46 \% DSC and 10.06 \% ASD (0.93 to 2.37 \% and 8.3 to 12.7 \%, respectively, across datasets) can be achieved with \ac{FastSurferVINN}}.

Overall, the inherent voxel size independence of \ac{FastSurferVINN} and resulting multi-resolution training option is highly beneficial to both, submillimeter and \SI{1.0}{\milli\meter} scans. 

\subsection{Big-FastSurferVINN}\label{sec:big}

Based on the observed performance gain with a larger training corpus, we evaluate whether \ac{ASD} and \ac{DSC} can be further improved across resolutions by expanding the training set from 120 to 1315 cases when exclusively adding  \SI{1.0}{\milli\meter} scans.

\begin{figure}[!hbt]
    \centering
    \includegraphics[width=\columnwidth,keepaspectratio]{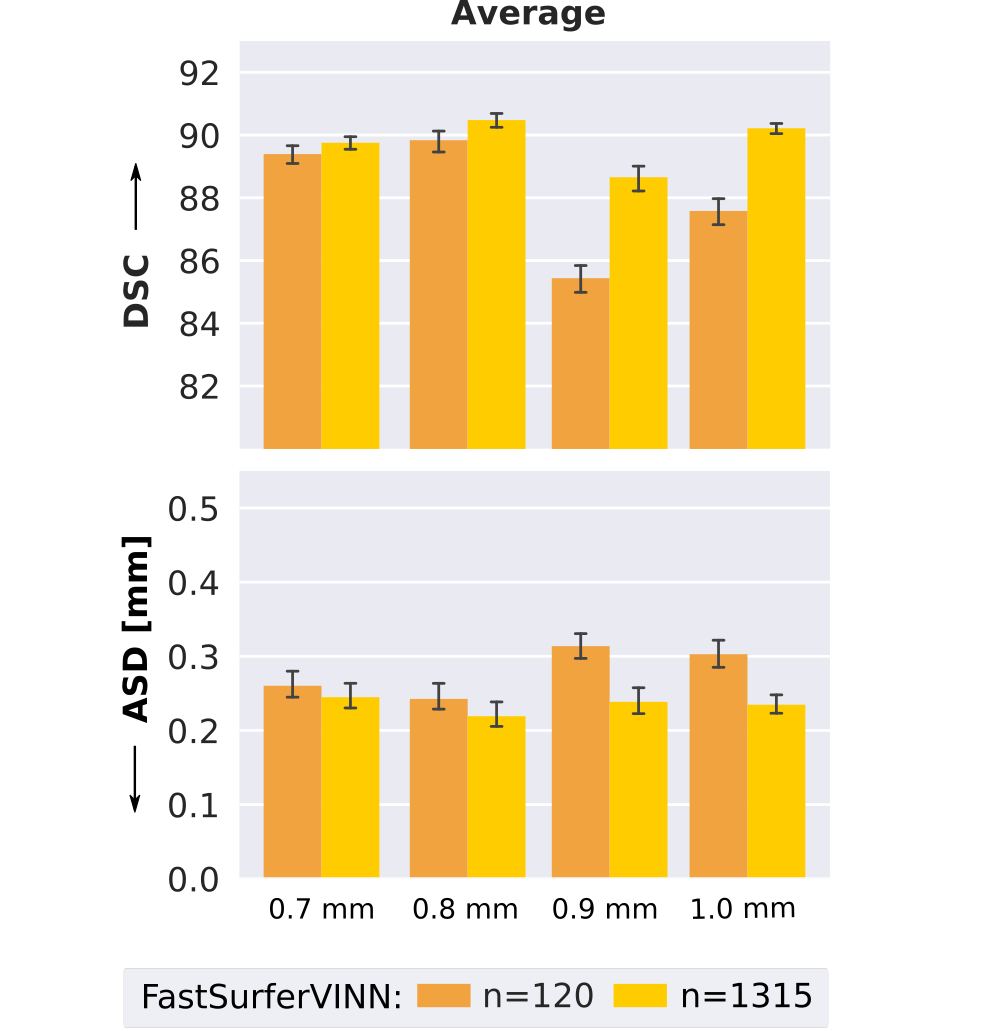}
    \caption{Big- \ac{FastSurferVINN} trained with approximately 20 times more \SI{1.0}{\milli\meter} scans (n=1315, yellow) than the original version (n=120, orange) raises segmentation performance across resolutions. Dice Similarity Coefficient (DSC, top) and average surface distance (ASD, bottom) improve on the submillimeter (\SI{0.7}{\milli\meter} - \SI{0.9}{\milli\meter}) as well as \SI{1.0}{\milli\meter} scans. }
    \label{fig:big}
\end{figure}

As illustrated in \Cref{fig:big}, expansion of the training corpus (n=1315, yellow bar) leads to a noticeable performance gain. Specifically the \SI{1.0}{\milli\meter} dataset benefits from the sample increase with a final \ac{DSC} of 90.26 and an \ac{ASD} of \SI{0.231}{\milli\meter}. On average, a 2.95\% increase in the \ac{DSC} and 16.26\% decrease in the \ac{ASD} can be observed for the \SI{1.0}{\milli\meter} scans compared to the smaller training set (n=120, orange bar), representing a significant performance gain (corrected $p<10^{-20}$). Additionally, \ac{ASD} and \ac{DSC} are significantly improved for the \SI{0.7}{\milli\meter}, \SI{0.8}{\milli\meter} and most strongly for the \SI{0.9}{\milli\meter} scans ($p<10^{-5}$, Wilcoxon signed-rank test). Here, \ac{DSC} and \ac{ASD} improve by 3.77\% and 24.08\%, respectively. Note, that the \SI{0.9}{\milli\meter} dataset is exclusively present in the testing corpus and thus completely new to the networks. A significant resolution bias in the training set, hence, does not decrease, but rather elevate performance across \SI{1.0}{\milli\meter} and submillimeter scans. 

\section{Discussion}

In this paper, we present the first multi-resolution deep learning tool for accurate and efficient (sub)millimeter 3T MRI whole brain segmentation. \ac{FastSurferVINN} addresses the two main difficulties associated with \ac{HiRes} \acp{MRI}, namely limited availability and diversity specifically for certain groups (e.g.\ neurodegeneration or scanner types) as well as resolution non-uniformity. Applicability of deep learning approaches is generally restricted to domains were enough training data exists (traditionally \SI{1.0}{\milli\meter} scans). \ac{FastSurferVINN}s' network-integrated resolution-normalization provides independence to the input voxel grid during both, training and inference, and favourably extends processing spans to resolutions not explicitly included in the training corpus. Further, both, \ac{LowRes} and \ac{HiRes} scans, can be included during training and, in turn, benefit from each others favourable properties, i.e.\ coverage and size versus better representation of detailed structures. 

The current state-of-the-art to approximate voxel size independence in networks is data augmentation. Here, transformations with randomly sampled scale parameters are applied to the images (both intensity and label map) in the native space. In \ac{FastSurferVINN}, we shift this step into the latent space by replacing the first scale transition from a pooling/unpooling operation to interpolation-based down-/upsampling. This network-integrated resolution-normalization interpolates the continuous feature maps on both the encoder and decoder arms and avoids alteration of the underlying ground truth labels entirely. 

When optimizing the network architecture, we first show that the best results are achieved with a 3$\times$3 kernel and bilinear interpolation. Also, our \ac{HiRes} loss which focuses segmentation performance on \ac{PVE}-affected structures (deep sulci, thin white matter strands and \ac{GM}-\ac{CSF} boundary) improves \ac{DSC} and \ac{ASD} specifically for the cortical structures. Interestingly, the improvement is consistent across resolutions, indicating that the structural information learned from the \ac{HiRes} scans is effectively transferred to the \SI{1.0}{\milli\meter} scans. Finally, we evaluate the introduction of attention mechanisms into the architecture. Overall, performance does not improve if the total number of parameters of the network with and without attention is controlled, indicating that network capacity (i.e.\ number of learnable parameters) rather than adaptive attention is the important factor here. 

We demonstrate \ac{FastSurferVINN}'s superior performance compared to state-of-the-art data augmentation for the fast and detailed segmentation of whole brain \ac{MRI}. Our network-integrated resolution-normalization in combination with internal scale augmentation outperforms traditional scale augmentation in terms of accuracy by a significant margin both with respect to FreeSurfer and a manual standard. Across nine different datasets and four resolutions, including both defaced (\ac{HCP}) and full-head images, our network achieves the highest \ac{DSC} compared to FreeSurfer as a reference (88.38 on average), as well as the lowest \ac{ASD} (\SI{0.283}{\milli\meter} on average). In addition, \ac{FastSurferVINN} achieves the best results on the manually labeled \SI{1.0}{\milli\meter} Mindboggle\-101 dataset with a \ac{DSC} of 81.89 and an AVG HD of \SI{0.471}{\milli\meter} on the cortical structures. Correspondingly, \ac{WM}, \ac{GM}, and hippocampus are better segmented with \ac{FastSurferVINN} on six manually corrected scans from the submillimeter \ac{RS} dataset. 

One possible explanation for the consistently improved segmentation metrics with our \ac{VINN} compared to traditional scale augmentation is the circumvention of label interpolation. For discrete labels, \ac{NN} or majority voting have to be applied. These kernels are prone to remove important structural details and create jagged segmentation maps \cite{Parker1983,Thevenaz2009,Schaum_1993,Allebach2005}. This is underlined by the observation that segmentation performance deteriorates for \ac{FastSurferVINN} when external scale augmentation is added. Further, analysis of the interpolation methods during the network-integrated resolution-normalization also highlights the negative impact of the \ac{NN} interpolation kernels. While all other methods (area, bi-linear, and bi-cubic) performed equally well, performance drops for \ac{NN} kernels on average by approximately 2\% \ac{DSC} and 10\% \ac{ASD}. Overall, \ac{NN} seems to have a systematic negative effect on segmentation performance and should, hence, be avoided wherever possible. 

In addition, our resolution-normalization interpolates a feature map of vectors instead of a single scalar image slice. As each vector describes a neighborhood in the input image, the available contextual information is extended before the interpolation step and may subsequently support a smoother transition between resolutions. Further, our network-integrated resolution-normalization reduces the range of perceived anatomical size variation in the inner blocks, thereby liberating resources to focus on structural details at a common metric scale (structures have similar sizes after normalization). Finally, a skip-connection transfers the non-interpolated feature maps from the pre-IDB into the post-CDB, potentially improving ambiguous label borders with details available at the native resolution. The differentiation between native voxel-scale and normalized inner scale may specifically be helpful in settings with large inter-subject size similarities (e.g.\ head size is relatively stable). As such, our network-integrated resolution-normalization approach is not limited to the neuroimaging context, but can be expected to benefit segmentation performance in other domains.

\ac{FastSurferVINN} further demonstrates excellent generalization performance with respect to inter- and extrapolation, achieving the best results for a variety of unseen submillimeter resolutions excluded from the training corpus. Especially more detailed cortical structures targeted with \ac{HiRes} acquisition show a significant \ac{DSC} increase on the \SI{0.7}{\milli\meter}, \SI{0.8}{\milli\meter} and \SI{0.9}{\milli\meter} scans, respectively. Similarly, \ac{ASD} improved by 19.7\%, 10\% and 11\% highlighting specifically the improved extrapolation capabilities of \ac{FastSurferVINN} on the left-out \SI{0.7}{\milli\meter} dataset component compared to traditional scale augmentation. Given the lack of a de-facto standard in \ac{HiRes} datasets, consistent performance towards unseen resolutions is an important property. Specifically, new releases such as the upcoming \ac{HCP} disease studies\footnote{\url{https://humanconnectome.org/disease-studies}} may benefit from a flexible and validated method to process scans with good accuracy at their native submillimeter resolution. Conveniently, \ac{FastSurferVINN} avoids lossy image downsampling, as well as time intensive re-training, and simplifies re-validation. One future aspect to investigate, is the extend to which this generalization is effective. Unfortunately, with increasing resolutions, evaluation becomes virtually impossible due to (i)~unavailability of manual segmentations and (ii)~limited support by existing neuroimaging pipelines (i.e.\ validation is difficult here as well due to point (i)). \new{Our substitute evaluation on the downsampled, lower resolution images underlines that \ac{FastSurferVINN} robustly generalizes beyond the tested resolution range (see \Cref{sec:loo}, \Cref{fig:resolution_lr}). Specifically, differences to traditional data augmentation are magnified with a more than 16.8 \% increase in \ac{DSC} for the \SI{1.6}{\milli\meter} scans. The improved generalizability of \ac{FastSurferVINN} to under- or unrepresented resolutions especially indicates superior performance in real applications.} Furthermore, \ac{FastSurferVINN} was trained and evaluated on 3T scans only. Adoption to ultra high-field \ac{MRI} capable of producing resolutions below \SI{0.5}{\milli\meter} is an interesting future avenue, as direct transferability is limited due to the strong signal, contrast, and noise differences \cite{Kolk2013,Zwaag2016}.

Overall, \ac{FastSurferVINN} allows processing of both \ac{LowRes} and \ac{HiRes} data without reducing precision on either set. On the contrary, we demonstrated that \ac{FastSurferVINN} even outperforms a fixed-resolution \SI{1.0}{\milli\meter} network by a significant margin. As the only difference in the training corpus is the increased image resolution of 60 subjects, the gain in accuracy seems to be predominantly motivated by the structural details provided by the submillimeter scans.

Further, the intensity and demographical variety added by the \SI{1.0}{\milli\meter} scans promoted generalization performance to the \ac{HCP}L component when \ac{FastSurferVINN} was compared to a fixed-resolution \ac{RS} \ac{HiRes} network. This is of special interest in longitudinal settings, where training on images from only the first time point may introduce a bias towards healthier or younger brains, which can be mitigated by introducing appropriate \ac{LowRes} out-of-study cases.

Similarly, potential future variations of acquisition parameters within studies may negatively affect performance of fixed-resolution networks and, hence, would require their retraining.  

Comparison to fixed-resolution networks revealed another promising application of \ac{FastSurferVINN}.

As shown in \Cref{sec:dedicated}, \ac{FastSurferVINN}'s segmentation performance improves across resolutions even if only additional \SI{1.0}{\milli\meter} scans are included. Surprisingly, performance is dominated by the training corpus size and not significantly influenced by its heterogeneous resolution. On \ac{RS}, the performance was comparable for \ac{FastSurferVINN} when training with a mixed resolution corpus (120 total, with only 30 \SI{0.8}{\milli\meter} \ac{RS} cases) as opposed to a fixed-resolution network trained exclusively on 120 \SI{0.8}{\milli\meter} \ac{RS} scans (see \Cref{sec:dedicated}). This can be exploited in settings where training cases are scarce at a specific resolution and new acquisitions difficult (e.g.\ manual segmentations or custom sub-resolution acquisition protocols). FastSurfer-VINN’s resolution flexibility is optimally suited to integrate a breadth of already existing resources, reducing the amount of newly acquired data necessary to achieve good performance.

Finally, the results of Big-FastSurferVINN indicate that an imbalance between \SI{1.0}{\milli\meter} and submillimeter data distribution does not introduce a resolution bias.

On the contrary, increasing the \SI{1.0}{\milli\meter} component in Big-FastSurferVINN by a factor 20 improved segmentation performance across all resolutions (\SI{0.7} to \SI{1.0}{\milli\meter}). Specifically, the \SI{0.9}{\milli\meter} dataset (\ac{abide-ii}) benefits from the training set extension with an 3.67\% increase in \ac{DSC} and 22.11\% decrease in \ac{ASD}. Unlike \ac{HCP} and \ac{RS}, the \SI{0.9}{\milli\meter} \ac{abide-ii} scans were acquired on a Philips scanner, which is present in a larger proportion in the extended training set (30\% Philip scanners compared to 1.6\%). However, no submillimeter Philips scan was included at any point during training. The improved performance on \ac{abide-ii}, therefore, highlights the potential for \ac{FastSurferVINN} to actively reduce scanner-biases through inclusion of \SI{1.0}{\milli\meter} \acp{MRI}. Exploring an expected alleviation of age- or disease-biases in submillimeter datasets with \ac{FastSurferVINN} presents an interesting direction for future work.

Overall, we introduce a fast, voxel size independent neural network that scales well to large datasets and enables seamless integration of a variety of resolutions during both, training and inference. Thereby, \ac{FastSurferVINN} offers the potential to improve  generalization performance to future \ac{HiRes} datasets without retraining, reduce potentially existing dataset biases, and curtail necessary labour and time intensive manual labeling efforts. 

\ac{FastSurferVINN} will be made available as part of the open source FastSurfer \cite{Henschel_2020} package\footnote{\url{https://github.com/Deep-MI/FastSurfer}}.

\section{Acknowledgements}

This work was supported by DZNE institutional funds, by the Federal Ministry of Education and Research of Germany (031L0206, 01GQ1801), and by NIH (R01 LM012719, R01 AG064027, R56 MH121426, and P41 EB030006). 

We would like to thank the Rhineland Study group (PI Monique M.B.\ Breteler) for supporting the data acquisition and management.
We also thank Rika Etteldorf for providing corrected high-resolution labels.

Data used in the preparation of this article were obtained in part by the OASIS Cross-Sectional with principal investigators D. Marcus, R, Buckner, J, Csernansky J. Morris; P50 AG05681, P01 AG03991, P01 AG026276, R01 AG021910, P20 MH071616, U24 RR021382, and OASIS: Longitudinal: Principal Investigators: D. Marcus, R, Buckner, J. Csernansky, J. Morris; P50 AG05681, P01 AG03991, P01 AG026276, R01 AG021910, P20 MH071616, U24 RR021382. Further, data used in the preparation of this article were obtained from the MIRIAD database. The MIRIAD investigators did not participate in analysis or writing of this report. The MIRIAD dataset is made available through the support of the UK Alzheimer's Society (Grant RF116). The original data collection was funded through an unrestricted educational grant from GlaxoSmithKline (Grant 6GKC). 
Data used in preparation of this article were obtained from the Alzheimer’s Disease Neuroimaging Initiative (ADNI)  database  (\url{adni.loni.usc.edu}).  As  such,  the  investigators  within  the  ADNI  contributed  to  the  design and implementation of ADNI and/or provided data but did not participate in analysis or writing of this report. A complete listing of ADNI investigators can be found at: \url{http://adni.loni.usc.edu/wp-content/uploads/how\_to\_apply/ADNI\_Acknowledgement\_List.pdf}.
Data collection and sharing for this project was funded by the Alzheimer's Disease Neuroimaging Initiative (ADNI) (National Institutes of Health Grant U01 AG024904) and DOD ADNI (Department of Defense award number  W81XWH-12-2-0012). ADNI is funded by  the National  Institute  on  Aging,  the  National  Institute  of Biomedical  Imaging  and  Bioengineering,  and  through  generous  contributions  from  the  following:  AbbVie, Alzheimer’s Association; Alzheimer’s Drug Discovery Foundation; Araclon Biotech; BioClinica, Inc.; Biogen; Bristol-Myers Squibb Company; CereSpir, Inc.; Cogstate; Eisai Inc.; Elan Pharmaceuticals, Inc.; Eli Lilly and Company; EuroImmun; F. Hoffmann-La Roche Ltd and its affiliated company Genentech, Inc.; Fujirebio; GE Healthcare; IXICO  Ltd.; Janssen Alzheimer Immunotherapy Research  \&  Development, LLC.; Johnson \& Johnson Pharmaceutical  Research \& Development LLC.; Lumosity; Lundbeck; Merck  \&  Co., Inc.; Meso Scale Diagnostics, LLC.; NeuroRx Research;   Neurotrack Technologies; Novartis Pharmaceuticals Corporation; Pfizer Inc.; Piramal Imaging; Servier; Takeda Pharmaceutical Company; and Transition Therapeutics. The Canadian Institutes of Health Research is providing funds to support ADNI clinical sites in Canada. Private sector contributions are facilitated by the Foundation for the National Institutes of Health (www.fnih.org). The grantee  organization is the Northern California Institute for  Research  and  Education, and the study is coordinated by the Alzheimer’s Therapeutic Research Institute at the University of Southern California. ADNI data are disseminated by the  Laboratory for Neuro Imaging at the University of Southern California. Data were also provided in part by the Human Connectome Project, WU-Minn Consortium (Principal Investigators: David Van Essen and Kamil Ugurbil; 1U54MH091657) funded by the 16 NIH Institutes and Centers that support the NIH Blueprint for Neuroscience Research; and by the McDonnell Center for Systems Neuroscience at Washington University.

\bibliographystyle{elsarticle-num}
\bibliography{mybibfile}

\section*{Appendix}
\setcounter{table}{0}
\renewcommand\thetable{\Alph{table}}
\setcounter{equation}{0}
\renewcommand\theequation{\Alph{equation}}
\setcounter{section}{1}
\renewcommand\thesection{\Alph{section}}

\subsection{Attention}\label{sec:appendix_attention}
A context-driven learnable attention mechanism following the method proposed in \cite{Qin2018} is evaluated as a reference. This modular mechanism can be inserted into any CDB and offers variable placement across the network architecture. Here, we investigate the addition of attention in the pre-IDB and post-CDB (see \Cref{sec:fastsurfer}) as they are predominantly dealing with feature maps at the native image resolution. In short, a sequence of two convolutions ($\tilde{H}_1^a$:~3$\times$3 kernel, half the number of filters as the input; $\tilde{H}_2^a$:~1$\times$1 kernel, one more filter than convolutions in the dense blocks) is followed by a final element-wise softmax function normalizing the activations, denoted as $\lambda_0$, $\lambda_1$, $\lambda_2$ and $\lambda_3$, to a sum of 1. The activation map $\lambda_0$ is used for weighting the input features representing an effective receptive field of 1. Each voxel is, therefore, assigned a different attention value driven by the image context. In contrast to \cite{Qin2018}, the feature responses at different scales ($c_1$, $c_2$ and $X_a$) are generated by a sequence of convolutions within the CDB ($H_a^1$, $H_a^2$ and $H_a^3$). Each convolution increases the receptive field size by $(k_1-1 \times k_2-1)$ with $(k_1 \times k_2)$ representing the kernel size. The calculated attention maps are then used to weight the local skip connections which combine the outputs of each convolution within these blocks. This is mathematically formulated in \Cref{eq:1,eq:2,eq:3}.
\begin{subequations}
\begin{align}
    \mathbf{X_a} = \lambda_3 H_a^3(c_2) \label{eq:1} \\
    c_2 = max(\lambda_2 H_a^2(c_1), c_1) \label{eq:2}\\
    \underbrace{c_1 = max(\lambda_1 H_a^1(X_{a-1}) , \lambda_0 X_{a-1})}_\text{Attention CDB} \label{eq:3}
\end{align}
\end{subequations}

\subsection{Ablative Optimization of FastSurferCNN}\label{ssec:appendix_ablation}

Ablative evaluation of the CNN base architecture optimizations are summarized in \Cref{tab:fastsurfer_base_dsc,tab:fastsurfer_base_asd}. Training and validation are based on the multi-resolution  datasets listed in \Cref{tab:datasets} (\textit{Mix}). The original FastSurferCNN CDBs consecutively perform two 5$\times$5 convolution operations followed by a final 1$\times$1 convolution (\Cref{sec:fastsurfer}, first row in \Cref{tab:fastsurfer_base_dsc,tab:fastsurfer_base_asd}). In total, an average \ac{DSC} of 88.63 and 87.09  and a \ac{ASD} of \SI{0.317}{\milli\meter} and \SI{0.283}{\milli\meter} is reached for the subcortical and cortical structures, respectively. Changing the kernel size to 3$\times$3 while keeping the effective receptive field size of 9$\times$9 per CDB constant (\Cref{tab:fastsurfer_base_dsc,tab:fastsurfer_base_asd}, 3$\times$3, 64 F) leads to a significant improvement in \ac{DSC} (p\textless0.01, Wilcoxon-signed rank test). Increasing the number of filters per layer from 64 to 71 preserves the original number of trainable parameters (approximately $1.85 * 10^6$). This change leads to a slight improvement on the subcortical and cortical structures. For comparability to our final \ac{VINN}, we also add the \new{pre-IDB} and \new{post-CDB} blocks (\Cref{tab:fastsurfer_base_dsc,tab:fastsurfer_base_asd}, \mbox{FastSurferCNN${}^*$}). This addition merely assures comparability between the augmentation and interpolation approach. As visible in \Cref{tab:fastsurfer_base_dsc,tab:fastsurfer_base_asd}, this change improves segmentation accuracy further with a final \ac{DSC} of 88.85 and 88.01 and an \ac{ASD} of \SI{0.307}{\milli\meter} and \SI{0.257}{\milli\meter}. The optimized \mbox{FastSurferCNN${}^*$} architecture is used as the baseline for all comparisons (i.e. augmentation and interpolation). 

\begin{table}[h!]
\caption{FastSurferCNN optimization. Dice Similarity Coefficent evaluation. }
\label{tab:fastsurfer_base_dsc}
\resizebox{\columnwidth}{!}{%
\begin{tabular}{l|l|l|}
                        & \multicolumn{2}{c|}{\textbf{DSC}}                     \\
\textbf{Datasets}       & \textbf{Subcortical}     & \textbf{Cortical}          \\ \hline
FastSurferCNN           & 88.63  ($\pm$1.87)       & 87.09  ($\pm$2.4)          \\
\rowcolor[HTML]{EFEFEF} 
+ 3x3, 64F              & 88.75  ($\pm$1.78)       & 87.68  ($\pm$2.35)         \\
+ 3x3, 71F              & 88.79  ($\pm$1.79)       & 87.78  ($\pm$2.41)         \\
\rowcolor[HTML]{EFEFEF} 
\textbf{FastSurferCNN*} & \textbf{88.85  ($\pm$2)} & \textbf{88.01  ($\pm$2.4)}
\end{tabular}%
}
\end{table}

\begin{table}[h!]
\caption{FastSurferCNN optimization. Average Surface Distance evaluation. }
\label{tab:fastsurfer_base_asd}
\resizebox{\columnwidth}{!}{%
\begin{tabular}{l|l|l|}
                        & \multicolumn{2}{c|}{\textbf{ASD}}                           \\
\textbf{Datasets}       & \textbf{Subcortical}         & \textbf{Cortical}            \\ \hline
FastSurferCNN           & 0.317  ($\pm$0.081)          & 0.283  ($\pm$0.085)          \\
\rowcolor[HTML]{EFEFEF} 
+ 3x3, 64F              & 0.312  ($\pm$0.077)          & 0.267  ($\pm$0.08)           \\
+ 3x3, 71F              & 0.311  ($\pm$0.08)           & 0.264  ($\pm$0.084)          \\
\rowcolor[HTML]{EFEFEF} 
\textbf{FastSurferCNN*} & \textbf{0.307  ($\pm$0.077)} & \textbf{0.257  ($\pm$0.082)}
\end{tabular}%
}
\end{table}

\subsection{\SI{1.0}{\milli\meter} Datasets}\label{sec:appendix_datasets}

\vspace{0.0ex}
    \textbf{ABIDE I}: The Autism Brain Imaging Data Exchange I ~\cite{abide_i_dataset} is a cross-sectional study involving 17 international sites and focuses on autism spectrum disorders. It contains data for 1112 individuals between 7 and 64 years of age.  Scanner and sequence parameters vary depending on the site and can be accessed on the ABIDE website (\url{https://fcon_1000.projects.nitrc.org/indi/abide/abide_I.html}). All \ac{MRI} data were acquired using 3 Tesla scanners (either Philips, Siemens or GE). 20 cases from the ABIDE-I were used for testing, 68 for training Big-FastSurferVINN. 
   
\vspace{0.5ex}
    \textbf{ADNI}: The Alzheimer’s Disease Neuroimaging Initiative~\cite{ADNI_dataset} was launched in 2003 as a public-private partnership, led by principal investigator Michael W.\ Weiner, MD. The ADNI database has $>$ 2000 participants and is available online at \url{http://adni.loni.usc.edu}. The primary goal of ADNI has been to test whether serial \ac{MRI},  positron  emission  tomography,  other  biological  markers,  and  clinical  and neuropsychological assessment can be combined to measure the progression of mild cognitive impairment and early Alzheimer’s disease (see \url{www.adni-info.org} for up-to-date information). Data were acquired at a resolution of 1.0x1.0x1.2~mm with 1.5T and 3T-\acp{MRI} scanners from the three largest MRI vendors (GE, Philips and Siemens) using an MP-RAGE sequence. Scanner parameters are optimized for the different vendors (see \cite{ADNI_mri} for details). 15 cases from ADNI where used for training, 8 for validation and 215 for training Big-FastSurferVINN. 40 different cases were used for assessing accuracy and generalizability in the final testset.

\vspace{0.5ex}
    \textbf{IXI}: This data collection provides 600 \acp{MRI} from normal, healthy subjects. The data has been collected at three different sites in London, UK on one GE (1.5T) and two Philips (1.5T and 3T) scanners and is available online (\url{https://brain-development.org/ixi-dataset/}) under Creative Commons License BY-NC-ND 3.0 (\url{https://creativecommons.org/licenses/by-sa/3.0/legalcode}). Detailed sequence parameters are available at the given URL. 43 scans from IXI were used for testing and 400 for training Big-FastSurferVINN.
    
\vspace{0.5ex}
    \textbf{LA5c}: The cross-sectional UCLA Consortium for Neuropsychiatric Phenomics LA5c Study~\cite{LA5c_dataset} includes 142 individuals diagnosed with a neuropsychiatric or neurodevelopmental disorder (schizophrenia, bipolar disorder, ADHD) and 130 normal controls (ages 21-50). All participants were scanned on a 3T Siemens Trio at a single\--center. T1-weighted MP-RAGE images were acquired with field of view of 250, 256x256 matrix, and 176 \SI{1.0}{\milli\meter} sagittal partitions. An TI of \SI{1.1}{\second}, TE of 3.5-\SI{3.3}{\milli\second}, TR of \SI{2.53}{\second} and flip angle of 7$^{\circ}$ was used for all scans. This data was obtained from the OpenfMRI database (\url{https://openfmri.org/dataset/ds000030/}). Its accession number is ds000030. 16 cases from LA5c were used for training, 9 for validation, 203 for training Big-FastSurferVINN, and 15 for the final testset.
    
\vspace{0.5ex}
    \textbf{Mindboggle-101}: The largest manually corrected set of free, publicly accessible (\url{https://osf.io/nhtur/}) labeled brain images based on a consistent human cortical labeling protocol (DKTatlas) \cite{Klein2012}. Mindboggle-101 consists of 101 labeled brain surfaces and volumes derived from T1-weighted brain MRIs of healthy individuals. Except for five subjects (MMRR-3T7T-2, Twins-2, and Afterthought-1), all \acp{MRI} are from publicly available collections (i.e.\ Test-Retest OASIS1 \cite{oasis_1_dataset},  the Multi-Modal Reproducibility Resource \cite{Landman2011}, Nathan Kline Institute Test–Retest and Nathan Kline Institute/Rockland Sample, Human Language Network subjects \cite{Morgan2009}, and Colin Holmes 27 template \cite{Holmes1998}, see \cite{Klein2012} for details). Manually labeled subcortical segmentations are available for the OASIS1 Test-Retest portion (20 subjects) within Mindboggle-101 (released under Creative Commons License BY-NC-ND 4.0 (\url{http://creativecommons.org/licenses/by-nc-nd/4.0}) by Neuromorphometrics, Inc. (\url{http://Neuromorphometrics.com/})). All 78 volumes with isotropic voxel sizes are used to evaluate network performance with respect to a manual reference.
    
\vspace{0.5ex}
    \textbf{MIRIAD}: The Minimal Interval Resonance Imaging in Alzheimer's Disease~\cite{miriad_dataset} is a publicly available longitudinal study with focus on neurodegeneration (see \url{http://miriad.drc.ion.ucl.ac.uk/}). \acp{MRI} of 23 elderly controls and 46 Alzheimer's diseased patients (ages 55+) were acquired at a single center with a 1.5T Signa MRI scanner (GE Medical systems), using an inversion recovery prepared fast spoiled gradient recalled  sequence, field of view of \SI{24}{\centi\meter}, 256x256 matrix, 124 \SI{1.5}{\milli\meter} coronal partitions (voxel size 0.9x0.9x1.5), TR \SI{15}{\milli\second}, TE \SI{5.4}{\milli\second}, flip angle 15$^{\circ}$, and TI \SI{650}{\milli\second}. 7 cases from MIRIAD were used for training and validation, respectively, and 23 for training Big-FastSurferVINN. 

\vspace{0.5ex}
   \textbf{Mind-Brain-Body}: The MPI Leipzig Mind-Brain-Body cohort \cite{mind-brain-body_2019,mind-brain-body2_2019} consists of 321 healthy participants between 20 and 77 years of age. \acp{MRI} were acquired on a Siemens Verio 3T with a weighted T1 Magnetization-Prepared 2 Rapid Acquisition Gradient Echoes (MP2RAGE) protocol and sagittal acquisition orientation, one 3D volume with 176 slices, TR of \SI{5000}{\milli\second}, TE of \SI{2.92}{\milli\second}, TI1 of \SI{700}{\milli\second}, TI2 of \SI{2500}{\milli\second}, flip angle 1 or 4 degrees, flip angle 2 of 5 degrees, echo spacing of \SI{6.9}{\milli\second}, \SI{1.0}{\milli\meter} isotropic voxel size and field of view of \SI{256}{\milli\meter}. This data was obtained from the OpenfMRI database. Its accession number is ds000221. 195 \acp{MRI} were used for training Big-FastSurferVINN.

\vspace{0.5ex}
    \textbf{Oasis-1}~\cite{oasis_1_dataset} and \textbf{Oasis-2}~\cite{oasis_2_dataset}: The Open Access Series of Imaging Studies 1 and 2, are publicly available (\url{https://www.oasis-brains.org/}) cross-sectional (Oasis-1) and longitudinal (Oasis-1) studies covering non-demented and demented individuals with very mild to moderate Alzheimer's disease. All subjects were scanned at a single-center using either a 1.5T Vision or a 3T TIM Trio Siemens Scanner in sagittal orientation with a voxel resolution of 1.0x1.0x1.25~mm. For acquisition, a MP-RAGE sequence with a TR of \SI{9.7}{\milli\second}, TE of \SI{4.0}{\milli\second}, flip angle of 10$^{\circ}$ and TI of \SI{20}{\milli\second} was used. Oasis-1 includes 416 subject between ages 18 to 96, while the longitudinal Oasis-2 focuses on older adults (150 subjects at age 60+). 14 cases from Oasis-1 and 8 from Oasis-2 were used for training, 11 and 5 for validation, 79 and 65 for training Big-FastSurferVINN and 35 and 17 for final testing.

\vspace{1ex}
Participants of the individual studies gave informed consent in accordance with the Institutional Review Board at each of the participating sites. Complete ethic statements are available at the respective study webpages.

\begin{table*}[h!]
\centering
\caption{Summary of datasets used for training, validation and testing. Table lists the scanner manufacturer, field strength, (disease) groups, age range and isotropic resolution (Res) used in the paper.  }
\begin{tabular}{l|l|l|l|l|l}
\textbf{Dataset}  & \textbf{Scanner}   & \textbf{1.5T/3T} & \textbf{Groups}    & \textbf{Age} & \textbf{Res} \\ \hline
HCP               & Siemens            & 3T               & Normal            & 22-35        & \SI{0.7}{\milli\meter}     \\
HCPL              & Siemens            & 3T               & Normal            & 25-75        & \SI{0.8}{\milli\meter}    \\
RS                & Siemens            & 3T               & Normal            & 30-95        & \SI{0.8}{\milli\meter}     \\
ABIDE-II ETHZ1 & Philips            & 3T               & ASD/Normal        & 20-31        & \SI{0.9}{\milli\meter}    \\ \hline
ABIDE-I           & Philips/GE/Siemens & 3T               & ASD/Normal        & 18-64        & \SI{1.0}{\milli\meter}               \\
ADNI              & Philips/GE/Siemens & 1.5T/3T          & AD/MCI/Normal     & 55-93        & \SI{1.0}{\milli\meter}               \\
IXI               & Philips/GE         & 1.5T/3T          & Normal            & 19-87        & \SI{1.0}{\milli\meter}               \\
LA5C              & Siemens            & 3T               & Neuropsych/Normal & 21-50        & \SI{1.0}{\milli\meter}               \\
MBB               & Siemens            & 3T               & Normal            & 20-77        & \SI{1.0}{\milli\meter}               \\
MIRIAD            & GE                 & 1.5T             & AD/Normal         & 55-86        & \SI{1.0}{\milli\meter}               \\
OASIS1            & Siemens            & 1.5T/3T          & Normal            & 18-90        & \SI{1.0}{\milli\meter}               \\
OASIS2            & Siemens            & 1.5T/3T          & AD/Normal         & 60-96        & \SI{1.0}{\milli\meter}              
\end{tabular}
\label{tab:datasources}
\end{table*}

\begin{table*}[h!]
\centering
\caption{Composition of training, validation and testing set used throughout the paper. The asterisk ($^*$) indicates submillimeter datasets are resampled to \SI{1.0}{\milli\meter}.  }
\begin{tabular}{l|l|l|l}
\multicolumn{2}{l|}{\textbf{Usage}}                                                                        & \textbf{Datasets (Subjects)}                                                                                                                                                            & \textbf{n}               \\ \hline
                             & Mix                                            & \begin{tabular}[c]{@{}l@{}}HCP (30), RS (30), ADNI (15), LA5C (16), MIRIAD (7), \\ OASIS1 (14), OASIS2 (8)\end{tabular}                                         & 120  \\
                             & \cellcolor[HTML]{EFEFEF}No \SI{0.8}{\milli\meter}                                                              & \cellcolor[HTML]{EFEFEF}\begin{tabular}[c]{@{}l@{}}HCP (60), ADNI (15), LA5C (16), MIRIAD (7), \\ OASIS1 (14), OASIS2 (8)\end{tabular}                                                                          & \cellcolor[HTML]{EFEFEF}120                          \\
                             & No \SI{0.7}{\milli\meter}                                    & \begin{tabular}[c]{@{}l@{}}RS (60), ADNI (15), LA5C (16), MIRIAD (7), \\ OASIS1 (14), OASIS2 (8)\end{tabular}                                                   & 120  \\
                             & \cellcolor[HTML]{EFEFEF}Only \SI{0.8}{\milli\meter}                                                             & \cellcolor[HTML]{EFEFEF}RS (small=60, big=120)                                                                                                                                                                                 & \cellcolor[HTML]{EFEFEF}60/120                           \\
                             & Only \SI{1.0}{\milli\meter}                                                            & \begin{tabular}[c]{@{}l@{}}\mbox{HCP${}^*$} (30), \mbox{RS${}^*$} (30), ADNI (15), LA5C (16), MIRIAD (7), \\ OASIS1 (14), OASIS2 (8)\end{tabular}                                                               & 120                          \\
\multirow{-11}{*}{Training}   & \cellcolor[HTML]{EFEFEF}Mix (Big)                                           & \cellcolor[HTML]{EFEFEF}\begin{tabular}[c]{@{}l@{}}HCP (30), RS (30), ABIDE-I (68), ADNI (215), IXI (400), \\ LA5C (203), MBB (195), MIRIAD (30), OASIS1 (79), OASIS2 (65)\end{tabular} & \cellcolor[HTML]{EFEFEF}1315 \\ \hline
                             & \begin{tabular}[c]{@{}l@{}}Mix (Big), \\ No \SI{0.8}{\milli\meter}, \\ No \SI{0.7}{\milli\meter}\end{tabular} & \begin{tabular}[c]{@{}l@{}}HCP (20), RS (20), ADNI (8), LA5C (9), MIRIAD (7), \\ OASIS1 (11), OASIS2 (5)\end{tabular}                                                                   & 80                           \\
                             & \cellcolor[HTML]{EFEFEF}Only \SI{0.8}{\milli\meter}                                    & \cellcolor[HTML]{EFEFEF}RS (20)                                                                                                                                                         & \cellcolor[HTML]{EFEFEF}20   \\
\multirow{-5}{*}{Validation} & Only \SI{1.0}{\milli\meter}                                                             & \begin{tabular}[c]{@{}l@{}}\mbox{HCP${}^*$} (20), \mbox{RS${}^*$} (20), ADNI (8), LA5C (9), MIRIAD (7), \\ OASIS1 (11), OASIS2 (5)\end{tabular}                                                                 & 80                           \\ \hline
                             & \cellcolor[HTML]{EFEFEF}Mix                                           & \cellcolor[HTML]{EFEFEF}\begin{tabular}[c]{@{}l@{}}HCP (80), RS (80), ABIDE-II (25), ABIDE-I (20), ADNI (40), \\ IXI (43), LA5C (15), OASIS1 (30), OASIS2 (17)\end{tabular}             & \cellcolor[HTML]{EFEFEF}350     \\
                             & \begin{tabular}[c]{@{}l@{}}No \SI{0.7}{\milli\meter}, \\ No \SI{0.8}{\milli\meter} \end{tabular}                                                               & HCP (80), RS (80), ABIDE-II (25)                                                                                                                                                        & 185                          \\
                             & \cellcolor[HTML]{EFEFEF}Manual Labels                                       & \cellcolor[HTML]{EFEFEF}RS (6), Mindboggle (78)                                                                                                                                         & \cellcolor[HTML]{EFEFEF}84   \\
                             & \begin{tabular}[c]{@{}l@{}}Only \SI{0.8}{\milli\meter}, \\ Only \SI{1.0}{\milli\meter} \end{tabular}                                                                     & \begin{tabular}[c]{@{}l@{}}RS (102), HCPL (10), ABIDE-I (20), ADNI (40), IXI (43), \\ LA5C (15), OASIS1 (36), OASIS2 (17)\end{tabular}                                                   & 259                          \\
\multirow{-7}{*}{Testing}    & \cellcolor[HTML]{EFEFEF}Mix (Big)                                            & \cellcolor[HTML]{EFEFEF}\begin{tabular}[c]{@{}l@{}}HCP (80), RS (80), ABIDE-II (25), ABIDE-I (20), ADNI (40),\\  IXI (43), LA5C (15), OASIS1 (35), OASIS2 (17)\end{tabular}             & \cellcolor[HTML]{EFEFEF}355 
\end{tabular}
\label{tab:datasets}
\end{table*}

\begin{table*}[h!]
\centering
\caption{FastSurfer (FastS) internal segmentation IDs and mapping to FreeSurfer (FreeS).}
\begin{tabular}{llll|l|l|l|}
\cline{1-3} \cline{5-7}
\multicolumn{1}{|l|}{\textbf{Subcortical Structures}} & \multicolumn{1}{l|}{\textbf{FastS}} & \multicolumn{1}{l|}{\textbf{FreeS}} &  & \textbf{Cortical Structures}      & \textbf{FastS} & \textbf{FreeS} \\ \cline{1-3} \cline{5-7} 
\multicolumn{1}{|l|}{Cortical white matter (lh)}      & \multicolumn{1}{l|}{1}                 & \multicolumn{1}{l|}{2}                   &  & caudalanteriorcingulate (lh)      & 34                & 1002                \\
\multicolumn{1}{|l|}{Lateral Ventricle (lh)}          & \multicolumn{1}{l|}{2}                 & \multicolumn{1}{l|}{4}                   &  & caudalmiddlefrontal (lh, rh)      & 35                & 1003, 2003          \\
\multicolumn{1}{|l|}{Inferior Lateral Ventricle (lh)} & \multicolumn{1}{l|}{3}                 & \multicolumn{1}{l|}{5}                   &  & cuneus (lh)                       & 36                & 1005                \\
\multicolumn{1}{|l|}{Cerebellar White Matter (lh)}    & \multicolumn{1}{l|}{4}                 & \multicolumn{1}{l|}{7}                   &  & entorhinal (lh, rh)               & 37                & 1006, 2006          \\
\multicolumn{1}{|l|}{Cerebellar Cortex (lh)}          & \multicolumn{1}{l|}{5}                 & \multicolumn{1}{l|}{8}                   &  & fusiform (lh, rh)                 & 38                & 1007, 2007          \\
\multicolumn{1}{|l|}{Thalamus (lh)}                   & \multicolumn{1}{l|}{6}                 & \multicolumn{1}{l|}{10}                  &  & inferiorparietal (lh, rh)         & 39                & 1008, 2008          \\
\multicolumn{1}{|l|}{Caudate (lh)}                    & \multicolumn{1}{l|}{7}                 & \multicolumn{1}{l|}{11}                  &  & inferiortemporal (lh, rh)         & 40                & 1009, 2009          \\
\multicolumn{1}{|l|}{Putamen (lh)}                    & \multicolumn{1}{l|}{8}                 & \multicolumn{1}{l|}{12}                  &  & isthmuscingulate (lh)             & 41                & 1010                \\
\multicolumn{1}{|l|}{Pallidum (lh)}                   & \multicolumn{1}{l|}{9}                 & \multicolumn{1}{l|}{13}                  &  & lateraloccipital (lh, rh)         & 42                & 1011, 2011          \\
\multicolumn{1}{|l|}{3rd-Ventricle}                   & \multicolumn{1}{l|}{10}                & \multicolumn{1}{l|}{14}                  &  & lateralorbitofrontal (lh)         & 43                & 1012                \\
\multicolumn{1}{|l|}{4th-Ventricle}                   & \multicolumn{1}{l|}{11}                & \multicolumn{1}{l|}{15}                  &  & lingual (lh)                      & 44                & 1013                \\
\multicolumn{1}{|l|}{Brain Stem}                      & \multicolumn{1}{l|}{12}                & \multicolumn{1}{l|}{16}                  &  & medialorbitofrontal (lh)          & 45                & 1014                \\
\multicolumn{1}{|l|}{Hippocampus (lh)}                & \multicolumn{1}{l|}{13}                & \multicolumn{1}{l|}{17}                  &  & middletemporal (lh, rh)           & 46                & 1015, 2015          \\
\multicolumn{1}{|l|}{Amygdala (lh)}                   & \multicolumn{1}{l|}{14}                & \multicolumn{1}{l|}{18}                  &  & parahippocampal (lh)              & 47                & 1016                \\
\multicolumn{1}{|l|}{CSF}                             & \multicolumn{1}{l|}{15}                & \multicolumn{1}{l|}{24}                  &  & paracentral (lh)                  & 48                & 1017                \\
\multicolumn{1}{|l|}{Accumbens (lh)}                  & \multicolumn{1}{l|}{16}                & \multicolumn{1}{l|}{26}                  &  & parsopercularis (lh, rh)          & 49                & 1018, 2018          \\
\multicolumn{1}{|l|}{Ventral DC (lh)}                 & \multicolumn{1}{l|}{17}                & \multicolumn{1}{l|}{28}                  &  & parsorbitalis (lh, rh)            & 50                & 1019, 2019          \\
\multicolumn{1}{|l|}{Choroid Plexus (lh)}             & \multicolumn{1}{l|}{18}                & \multicolumn{1}{l|}{31}                  &  & parstriangularis (lh, rh)         & 51                & 1020, 2020          \\
\multicolumn{1}{|l|}{Cortical white matter (rh)}      & \multicolumn{1}{l|}{19}                & \multicolumn{1}{l|}{41}                  &  & pericalcarine (lh)                & 52                & 1021                \\
\multicolumn{1}{|l|}{Lateral Ventricle (rh)}          & \multicolumn{1}{l|}{20}                & \multicolumn{1}{l|}{43}                  &  & postcentral (lh)                  & 53                & 1022                \\
\multicolumn{1}{|l|}{Inferior Lateral Ventricle (rh)} & \multicolumn{1}{l|}{21}                & \multicolumn{1}{l|}{44}                  &  & posteriorcingulate (lh)           & 54                & 1023                \\
\multicolumn{1}{|l|}{Cerebellar White Matter (rh)}    & \multicolumn{1}{l|}{22}                & \multicolumn{1}{l|}{46}                  &  & precentral (lh)                   & 55                & 1024                \\
\multicolumn{1}{|l|}{Cerebellar Cortex (rh)}          & \multicolumn{1}{l|}{23}                & \multicolumn{1}{l|}{47}                  &  & precuneus (lh)                    & 56                & 1025                \\
\multicolumn{1}{|l|}{Thalamus (rh)}                   & \multicolumn{1}{l|}{24}                & \multicolumn{1}{l|}{49}                  &  & rostralanteriorcingulate (lh, rh) & 57                & 1026, 2026          \\
\multicolumn{1}{|l|}{Caudate (rh)}                    & \multicolumn{1}{l|}{25}                & \multicolumn{1}{l|}{50}                  &  & rostralmiddlefrontal (lh, rh)     & 58                & 1027, 2027          \\
\multicolumn{1}{|l|}{Putamen (rh)}                    & \multicolumn{1}{l|}{26}                & \multicolumn{1}{l|}{51}                  &  & superiorfrontal (lh)              & 59                & 1028                \\
\multicolumn{1}{|l|}{Pallidum (rh)}                   & \multicolumn{1}{l|}{27}                & \multicolumn{1}{l|}{52}                  &  & superiorparietal (lh, rh)         & 60                & 1029, 2029          \\
\multicolumn{1}{|l|}{Hippocampus (rh)}                & \multicolumn{1}{l|}{28}                & \multicolumn{1}{l|}{53}                  &  & superiortemporal (lh, rh)         & 61                & 1030, 2030          \\
\multicolumn{1}{|l|}{Amygdala (rh)}                   & \multicolumn{1}{l|}{29}                & \multicolumn{1}{l|}{54}                  &  & supramarginal (lh, rh)            & 62                & 1031, 2031          \\
\multicolumn{1}{|l|}{Accumbens (rh)}                  & \multicolumn{1}{l|}{30}                & \multicolumn{1}{l|}{58}                  &  & transversetemporal (lh, rh)       & 63                & 1034, 2034          \\
\multicolumn{1}{|l|}{Ventral DC (rh)}                 & \multicolumn{1}{l|}{31}                & \multicolumn{1}{l|}{60}                  &  & insula (lh, rh)                   & 64                & 1035, 2035          \\
\multicolumn{1}{|l|}{Choroid Plexus (rh)}             & \multicolumn{1}{l|}{32}                & \multicolumn{1}{l|}{63}                  &  & caudalanteriorcingulate (rh)      & 65                & 2002                \\
\multicolumn{1}{|l|}{WM-hypointensities}              & \multicolumn{1}{l|}{33}                & \multicolumn{1}{l|}{77}                  &  & cuneus (rh)                       & 66                & 2005                \\ \cline{1-3}
                                                      &                                        &                                          &  & isthmuscingulate (rh)             & 67                & 2010                \\
                                                      &                                        &                                          &  & lateralorbitofrontal (rh)         & 68                & 2012                \\
                                                      &                                        &                                          &  & lingual (rh)                      & 69                & 2013                \\
                                                      &                                        &                                          &  & medialorbitofrontal (rh)          & 70                & 2014                \\
                                                      &                                        &                                          &  & parahippocampal (rh)              & 71                & 2016                \\
                                                      &                                        &                                          &  & paracentral (rh)                  & 72                & 2017                \\
                                                      &                                        &                                          &  & pericalcarine (rh)                & 73                & 2021                \\
                                                      &                                        &                                          &  & postcentral (rh)                  & 74                & 2022                \\
                                                      &                                        &                                          &  & posteriorcingulate (rh)           & 75                & 2023                \\
                                                      &                                        &                                          &  & precentral (rh)                   & 76                & 2024                \\
                                                      &                                        &                                          &  & precuneus (rh)                    & 77                & 2025                \\
                                                      &                                        &                                          &  & superiorfrontal (rh)              & 78                & 2028                \\ \cline{5-7} 
\end{tabular}
\label{tab:labels}
\end{table*}

\end{document}